\documentclass[12pt]{article}
\usepackage{epsfig,amsmath,amssymb,cite}

\numberwithin{equation}{section}

\begin{document}
\title{\begin{flushright}\normalsize
  CERN--TH/2000--189\\
  LPTHE--01--07\\
  hep-ph/0102343 \\
  February 2001\\
\end{flushright}\vspace{1cm}
\textbf{Hadron masses and power corrections to event
  shapes}\footnote{Work supported in part by the EU Fourth Framework
    Programme `Training and Mobility of Researchers', Network `Quantum
    Chromodynamics and the Deep Structure of Elementary Particles',
    contract FMRX-CT98-0194 (DG 12-MIHT).}}

\author{ \textbf{G.P. Salam$^{1,2}$ and D. Wicke$^3$}\\
  {\normalsize $^1$ TH Division, CERN, CH-1211 Gen\`eve 23}\\
   {\normalsize $^2$ LPTHE, Universit\'es P. \& M. Curie (Paris VI) et
     Denis Diderot (Paris VII), Paris, France
 }\\
  {\normalsize $^3$ EP Division, CERN, CH-1211 Gen\`eve 23}
}

\date{}

\maketitle

\begin{abstract}
  It is widely believed that hadronisation leads to $1/Q$ corrections
  to $e^+e^-$ event shapes. We show that there are further
  corrections, proportional to $(\ln Q)^A/Q$ with $A= 4C_A/\beta_0
  \simeq 1.6$, associated with hadron masses and whose relative
  normalisations can be calculated from one observable to another. At
  today's energies these extra corrections can be of the same order of
  magnitude as `traditional' $1/Q$ corrections. They fall into two
  classes: universal and non-universal. The latter can be eliminated
  by suitable redefinitions of the observables.
\end{abstract}
\thispagestyle{empty}

\newpage
~\thispagestyle{empty}
\newpage

\setcounter{page}{1}

%
%
%
%
%
%
%

\newcommand{\as}{\alpha_{\mathrm{s}}}
\def\cO#1{{\cal O}\left(#1\right)}
\def\order#1{{\cal O}\left(#1\right)}
\newcommand{\ee}{e^+e^-}
\newcommand{\etabar}{{\bar \eta}}
\newcommand{\qbar}{{\bar q}}
\newcommand{\cV}{{\cal V}}
\newcommand{\dVmass}{\langle \delta_{m} \cV \rangle}
\def\dmass#1{\langle \delta_{m} #1 \rangle}
\newcommand{\eff}{\mathrm{eff}}
\newcommand{\decay}{\mathrm{decay}}
\newcommand{\hadr}{\mathrm{hadr}}
\newcommand{\res}{\mathrm{res}}
\newcommand{\had}{\mathrm{had}}
\newcommand{\cM}{{\cal M}}
\newcommand{\cN}{{\cal N}}
\newcommand{\cE}{{\cal E}}
\newcommand{\NP}{\mathrm{NP}}
\newcommand{\cf}{\textit{cf.}\ }
\newcommand{\ie}{\textit{i.e.}\ }
\newcommand{\eg}{\textit{e.g.}\ }
\newcommand{\pscheme}{p\textrm{-scheme}}
\newcommand{\bE}{\mathbf{E}}
\newcommand{\bK}{\mathbf{K}}
\newcommand{\CA}{C_\mathrm{A}}
\newcommand{\CF}{C_\mathrm{F}}
\newcommand{\nf}{n_\mathrm{f}}
\newcommand{\om}{\omega}
\newcommand{\ga}{\gamma}

\newcommand{\allpower}{Webber94,Webber94tube,DW,DMW,DasW,KS,AZ,BB,NS,BBM,Milan,Milan2,DasWMilan,DMSmye,Broad,GG,Zoltan,Korchemsky}
\newcommand{\exppower}{exppoweree,exppowerDIS,Delphi,HERApscheme}
\newcommand{\allmilan}{BBM,Milan,Milan2,DasWMilan,DMSmye}
\newcommand{\alldata}{EEResults,exppoweree,Delphi}

\newcommand{\TeV}{\,\mbox{Te\kern-0.2exV}}
\newcommand{\GeV}{\,\mbox{Ge\kern-0.2exV}}
\newcommand{\MeV}{\,\mbox{Me\kern-0.2exV}}
\newcommand{\keV}{\,\mbox{ke\kern-0.2exV}}
\newcommand{\eV}{\,\mbox{e\kern-0.2exV}}

\newlength{\signlength}
\newcommand{\nosign}{\settowidth{\signlength}{$-$}\hspace*{\signlength}}

\section{Introduction
}

For many non-perturbative aspects of QCD, there exist well defined
methods of study such as lattice calculations or chiral perturbation
theory. But one of the most fundamental aspects of non-perturbative
QCD at high energy colliders, namely hadronisation, the relation
between what is calculated perturbatively (parton level) and what is
actually measured (hadron level) remains well beyond the reach of such
methods. At best there are complex models which appear to reproduce
much of the experimental data, but they leave something to be desired
in terms of a fundamental understanding of what is involved in
hadronisation.

Over the past few years there has been considerable theoretical
\cite{\allpower} and experimental \cite{\exppower} interest in the
study of hadronisation contributions to $\ee$ and DIS event shapes.
These variables provide a convenient `laboratory' for such studies
because the hadronisation effects are responsible for a significant
fraction of the observable, making it feasible to carry out
quantitative tests of the theoretical predictions.

A number of the current theoretical approaches are based on the idea
of extending the reach of perturbation theory to very low scales ---
thus hadronisation effects are argued to be related to the infrared
behaviour of the coupling, or equivalently to the high-order behaviour
of the perturbative series. All the methods predict that the
hadronisation (or power) corrections to event shapes should scale as
$1/Q$, where $Q$ is the hard scale of the process. They also aim to
predict the relative normalisation of the corrections from one
observable to another.

As a result of their perturbative origins, the theoretical
calculations are usually carried out with the assumption that all
particles are massless.  But in practice the observed hadrons do have
masses.  So in this paper we examine how the treatment of hadron
masses modifies one's expectations about power corrections.

It is perhaps worth illustrating how masses can affect our observables
with a simple example. Let us consider the event-shape variable $\rho$,
\ie the squared invariant jet mass, normalised to $Q^2$ (for brevity
we often refer to it simply as the jet mass).  Initially one might
consider a Born event consisting of a pair of back-to-back particles,
each of mass $m$. The jet mass has the value $\rho=m^2/Q^2$. If
this were the end of the story then we would argue that particle
masses can be neglected since they give a $1/Q^2$ correction.

But one also needs to consider events containing soft particles.  Here
the situation is qualitatively different: the jet mass is
$(E^2-p^2)/Q^2$, where $E$ and $p$ are the total jet energy and
$3$-momentum respectively. For a jet aligned along the $z$-axis a
particle with mass $m_i$, $z$-momentum $p_{zi}$ and energy $E_i$
contributes an amount $(E_i-p_{zi})/Q$ to the jet mass and this
difference includes a piece of order $m_i^2/(2E_i Q)$. For a soft
particle with $E_i \sim \Lambda$ this translates to a $1/Q$
contribution. So parametrically at least, mass-related effects from
soft particles are of the same order as traditional power corrections,
and should not be neglected.

This is not the end of the story, for the simple reason that an event
generally contains many soft particles: there will be a
multiplicity-related enhancement of the $m^2/Q E_i$ correction.  The
sum, over all particles, of $1/E_i$ is proportional to the
$(-1)^{\mathrm{th}}$ moment of particle energy fractions.
It has been known for a long time that positive
moments and even the zeroth moment of particle energy fractions
undergo logarithmic scaling violations with perturbatively calculable
anomalous dimensions. It turns out that such analyses can be extended
to negative moments, with the result that the sum $\sum_i 1/E_i$
scales as $(\ln Q)^A$, where $A = 4\CA/\beta_0\simeq1.6$, with
$\beta_0 = (11\CA - 2\nf)/3$. This means that the formally $1/Q$
contribution from mass effects is enhanced by a factor $(\ln Q)^A$.

The jet mass is an example of a variable in which it is quite
straightforward to see that there are mass effects; in other cases the
mass dependence can arise more subtly. For a general variable, the
ability to factorise transverse and longitudinal degrees of freedom
(with respect to the quark-antiquark directions) is an essential
element of traditional, `perturbative' approaches to hadronisation
corrections.  We will discover that for an ensemble of massive
particles, differences arise between the factorisations applying to
the event-shapes on the one hand and particle production on the other
--- and by studying this mismatch we become sensitive to hadron mass
effects.
  
We will also see that a given event-shape variable can be defined in
variety of ways (schemes) which are all equivalent for an ensemble of
massless particles, but differ if there are massive particles. Two
examples are a definition in terms of just $3$-momenta ($p$-scheme),
and a definition in terms of energies and angles ($E$-scheme). It
turns out that of these schemes, one, the $E$-scheme, is privileged,
because it has the property that there is no mismatch between the
event-shape and particle-production factorisation properties, even for
massive ensembles.

Does this mean that in the $E$-scheme there are no mass-dependent
contributions? The answer is no: it just means that any mass-dependent
contribution is proportional to the same coefficient $c_\cV$ as the
`traditional' non-perturbative correction. One can demonstrate that
there really are still mass-dependent corrections by studying the
difference between an event shape before hadrons have had the time to
decay and after they have decayed.

We shall present the different elements of our analysis as follows. In
section~\ref{sec:EvShapes} we define the event shapes that will be
considered and then quote their soft limits, being careful to leave in
the leading dependence on particle masses.

Then in section~\ref{sec:tube} we review the tube model for
non-perturbative effects \cite{FeynmanTube,Webber94tube}, which is
known to reproduce all basic predictions for universal power
corrections.  However in contrast to previous treatments we leave in
mass effects and see how they lead to non-universal $1/Q$ power
corrections. This lays the ground for introducing alternative
definitions of event shapes (massive, $p$, $E$ and decay-scheme
definitions) which are identical for massless ensembles but differ in
the treatment of particle masses.

Whereas the tube model is adequate for describing normal $1/Q$ power
corrections it cannot address issues such as the scaling violations of
energy moments which must be considered in order to derive the full
$Q$-dependence of the mass-dependent corrections. For this we need to
recall how coherent branching \cite{coherence} affects hadron
multiplicities and relate this to our particular problem. This is done
in section~\ref{sec:QCD}.

Then in section~\ref{sec:MC} we start to study some of the practical
aspects of mass corrections --- for example we examine how,
numerically, they compare to other contributions. We discuss some of
the issues related to the experimental measurement of mass effects (a
fairly difficult task) and we compare our predictions to results from
Monte Carlo event generators. There is good agreement with Herwig
\cite{Herwig} and Ariadne \cite{Ariadne}, but significant disagreement
with Pythia \cite{Pythia} concerning the energy dependence of mass
effects at very high energies.

Finally in section~\ref{sec:data} we use a Monte Carlo event-generator
to correct data for event shapes to a variety of schemes and see how
this affects the fitted values for the perturbative and
non-perturbative parameters $\as$ and $\alpha_0$. We also investigate
the feasibility of carrying out fits with an extra parameter intended
to allow a separation of hadronisation into `traditional'
hadronisation effects and mass-related effects.

Our conclusions \cite{GBU} are presented in section~\ref{sec:concl}.
The appendices contain a summary of the notation used and introduced
in this article (appendix~\ref{sec:notation}), some calculational
details related to the jet broadenings (appendix~\ref{sec:broad}),
some considerations about $1/Q^2$ power corrections associated with
heavy quark decay (appendix~\ref{sec:heavyquarks}), and a Monte-Carlo
study of some peculiar features of the heavy-jet mass
(appendix~\ref{sec:mh}).

\section{Event shapes}
\label{sec:EvShapes}

The basic event shapes that we shall consider are the thrust $T$, the
invariant jet mass $\rho$, the $C$-parameter and the total jet
broadening $B_T$. For an ensemble of particles with momenta $k_i$ they
are defined as follows:
\begin{subequations}
\label{eq:ShapeDefs}
\begin{align}
\label{eq:T}  
T &= \max_{\vec n} \frac{ \sum_i \vec k_i . \vec n}{\sum_i
  |{\vec k}_i|}\,,\\
\label{eq:rho}
\rho &= \frac{\left(\sum_i k_i \,\Theta(\vec k_i . \vec n_T) \right)^2}{
  \left(\sum_i    k_{0i}\right)^2}\,.\\
\label{eq:C}
C &= \frac32 \frac{\sum_{i,j} |{\vec k}_i| |{\vec k}_j| \sin^2
  \theta_{ij}}{\left(\sum_i    |{\vec k}_i|\right)^2}\,,\\
\label{eq:BT}
B_T &= \frac{\sum_i |{\vec k}_{ti}| }{ 2\sum_i |{\vec k}_i|}\,,
\end{align}
\end{subequations}
where $\theta_{ij}$ is the angle between particles $i$ and $j$,
$k_i=(k_{0i},\vec{k}_i)$ the four-momentum, and
${\vec k}_{ti}$ the transverse momentum of particle $i$ with
respect to the thrust axis $\vec n_T$.

We shall also discuss the heavy-jet mass $\rho_h$ which measures the
invariant jet mass in the heavier of the two hemispheres (separated by
the plane perpendicular to the thrust axis), and the wide-jet
broadening $B_W$, which measures the jet-broadening in the wider of
the two hemispheres.

For the purpose of calculating power corrections we will be interested
in the behaviour of the event shapes variables for two-jet
configurations, since it is such configurations which are the most
frequent, representing a fraction $1-\order{\as}$ of all events.  In
the two-jet limit all variables tend to zero, except the thrust which
tends to $1$. Accordingly it will be more convenient to refer to $\tau
= 1-T$.

Essentially a two-jet event consists of a pair of back-to-back hard
particles (which may have fragmented collinearly) and a bunch of accompanying
soft particles.  The hard particles (associated with the $q\qbar$ pair
at parton level) define the axis of the event, while the soft
particles (gluons at parton level) give the deviation of the event
shape variable from its Born value of zero.  Variables like $\tau$,
$\rho$ and $C$ are particularly simple (linear) in that, in the
two-jet limit their value is given by the sum of independent
contributions from each soft particle:
\begin{subequations}
\label{eq:ShapeDefsSoft}
\begin{align}
  \label{eq:tauSoft}
  \tau &\simeq \frac{1}{Q} \sum_{i \in \mathrm{soft}} k_{ti}\,
  e^{-|\etabar_i|} \,,\\
  \label{eq:rhoSoft}
  \rho &\simeq 
  \frac{1}{Q} \sum_{i \in \mathrm{soft}} \sqrt{k_{ti}^2 +
              m_i^2} \,e^{-\eta_i}\, \Theta(\eta_i)\,,
 \\ 
  C &\simeq \frac{1}{Q} \sum_{i \in \mathrm{soft}} \frac{3\,
    k_{ti}}{\cosh \etabar_i} \,,
\end{align}
where $\etabar_i$ is the pseudorapidity of particle $i$ with respect
to the thrust ($z$) axis, $\etabar_i = - \ln \tan \theta_i/2$ and
$\eta_i$ is its rapidity $\eta_i = \frac12 \ln \frac{k_{0i} + k_{zi}}{k_{0i}
  - k_{zi}}$.  For massless particles the pseudorapidity and the
rapidity are identical.  In these expressions we have neglected mass
effects in the denominators (in the case of $\tau$, only after going
from $T$ to $\tau$), since on average the numerator is small,
$\order{\as}$, and thus modifications of $\cO{\Lambda}$ to the
denominator give effects of $\order{\as \Lambda/Q}$, which we can
neglect.\label{para:suppressed_denominator}

Other variables like the heavy-jet mass and the broadenings are more
complex. To see why, let us consider the case of the heavy-jet mass:
when the event contains exactly one soft particle the heavy-hemisphere
is always that in which the soft particle is present and the soft
particle \emph{always} contributes to the heavy jet mass. But if there
are many soft particles then the hardest one (specifically, the one
with the largest $k_{ti}\, e^{-|\etabar_i|}$) determines which
hemisphere is heavy and any given softer particle contributes when it
is in the heavy hemisphere \ie only \emph{half} the time.  Thus the
heavy jet mass is not equal to a linear combination of independent
contributions from each soft particle.

For the purposes of studying non-perturbative corrections we can
however make a simplifying approximation: the `hardest' soft particles
come from perturbative emissions, while non-perturbative emissions
will be much softer. The contributions from these softest particles do
combine linearly \cite{AZ,Milan2}, so that one can write
\begin{equation}
  \label{eq:rhoHSoft}
  \rho_h \simeq 
  \frac{1}{Q} \sum_{i \in \mathrm{softest}} \sqrt{k_{ti}^2 +
              m_i^2} \,e^{-\eta_i}\, \Theta(\eta_i)  + \ldots\,,
\end{equation}
where we have taken the hard hemisphere as being that with $\eta
>0$ and the dots indicate the contribution from harder particles. For
both jet broadenings analogous arguments apply \cite{Milan2,Broad} and
one has
\begin{align}
  \label{eq:BTSoft}
  B_T &\simeq \frac{1}{2Q} \sum_{i \in \mathrm{softest}} k_{ti}\,
  + \ldots\\
  \label{eq:BWSoft}
  B_W &\simeq \frac{1}{2Q} \sum_{i \in \mathrm{softest}} k_{ti}\,
    \Theta(\etabar_i) + \ldots\,,
\end{align}
\end{subequations}
where again we have taken the wide hemisphere as being that with
$\etabar > 0$.

\section{Power corrections: the tube model}
\label{sec:tube}

A variety of approaches exist for the study of power corrections in
event shapes \cite{\allpower}.  The simplest, which reproduces the
results of the more sophisticated methods \cite{\allmilan}, is the
tube, or longitudinal phase-space model
\cite{FeynmanTube,Webber94tube}.

The principle behind the tube model (the ideas of \cite{KS} are
analogous, though presented in a more formal language) is as follows.
Soft hadrons (\ie the hadrons responsible for non-perturbative
corrections) are generated from the $q\qbar$ pair of sources; since
both sources are fast moving (in opposite directions) a moderate boost
along the $q\qbar$ direction still leaves us with two fast-moving
particles, and so does not change the structure of the low-transverse
momentum fields at central rapidities. As a result soft particle
production at central rapidities must be boost-independent.  Since a
boost just corresponds to a shift in rapidity, this is equivalent to
saying that non-perturbative (and in general soft) particle production
is rapidity-independent,  at least for rapidities $\eta \ll \ln Q/k_t$,
beyond which one becomes sensitive to the finite energy of the
source. The tube model makes no statement about the
transverse-momentum distribution of soft particles, so we just write
the distribution of non-perturbatively
produced\footnote{`Non-perturbatively produced' is a rather awkward
  term --- in section~\ref{sec:QCD} we examine in detail what we
  mean.} hadrons of type $h$ as being:
\begin{equation}\label{eq:tubedist}
  \frac{dn_h}{d\eta\, d \ln k_t} = \phi_h(k_t), 
    \qquad |\eta|\ll \ln \frac{Q}{k_t}\,,
\end{equation}
with $\phi_h(k_t)$ some a priori unknown function. There is no
$Q$-dependence in $\phi_h(k_t)$ since the fields generated by a source
($q$ or $\qbar$) moving close to the speed of light are independent of
the energy of the source.

\subsection{Massless case}
\label{sec:massless}

Most power correction analyses work within the approximation that
particles are massless.\footnote{Even the `massive-gluon' calculations
  make this approximation, since they generally assume that the
  massive gluon decays into two massless particles. An exception is to
  be found in \cite{BB}, as discussed briefly in
  section~\ref{sec:massivegluoncalc}.} As a result in
eqs.~\eqref{eq:ShapeDefsSoft} all explicit mass-dependence disappears,
and one can replace $\etabar$ with $\eta$. This leaves the expressions
for the event-shape variable $\cV$ in a \emph{factorised} form,
whereby all rapidity dependence (which differs from one observable to
another) can be separated from the transverse momentum dependence
($k_t/Q$ in all observables):
\begin{equation}
  \label{eq:linearity}
  \cV = \sum_{i \in \mathrm{soft}} \frac{k_{ti}}{Q} f_\cV(\eta_i)\,.
\end{equation}
For example for the the thrust we have $f_\tau(\eta) = e^{-|\eta|}$.
The non-perturbative contribution to the mean value of the event-shape
is then given by
\begin{equation}
  \label{eq:NPmassless}
  \langle \cV_{\NP} \rangle =  \int
  \frac{dk_t}{k_t} \,
  \frac{k_t}{Q} \sum_{h} \phi_h(k_t)  \int d\eta f_{\cV}(\eta) \;=\;
  \frac{a_0}{Q}\, c_{\cV} 
\end{equation}
where we have defined a non-perturbative parameter
\begin{equation}
  \label{eq:a0}
  a_0 = \int d k_t
  \sum_{h} \phi_h(k_t)
\end{equation}
and a calculable, variable-dependent coefficient,
\begin{equation}
  \label{eq:cV}
  c_{\cV} = \int d\eta\, f_{\cV}(\eta) \,.
\end{equation}
The factorisation of rapidity and $k_t$ dependence is the prerequisite
for \emph{universality}, namely the fact that the power corrections
for a range of observables all depend on the same, universal,
non-perturbative quantity ($a_0$), with a calculable coefficient ($c_\cV$). The
predictions for the coefficients are given in table~\ref{tab:cvs}.
\begin{table}[b]
\begin{center}
  \begin{tabular}{|c||c|c|c|c|c|c|}
\hline
$\cV$ & $\tau$ & $\rho$ & $\rho_h$ & $C$  & $B_T$ & $B_W$    \\ \hline
$c_\cV$ & $2$  &  $1$   &   $1$    & $3\pi$ & 
 $\frac{\pi}{2\sqrt{C_F \as}} - \frac{\beta_0}{6\CF} + \eta_0$ 
& $\frac{\pi}{4\sqrt{2C_F \as}}- \frac{\beta_0}{24\CF} + \frac{\eta_0}{2}$
\\ \hline
  \end{tabular}
\end{center}
  \caption{coefficients of $1/Q$ power corrections; $\beta_0
    =\frac{11}{3}\CA -\frac23\nf$ and $\eta_0 \simeq 0.13629$. }
\label{tab:cvs}
\end{table}

The more complex form for the broadenings \cite{Broad} arises because
these variables are sensitive to the mismatch between the thrust axis
and the quark axis. Emissions are uniform in rapidity with respect to
the latter, whereas the broadening measures $k_t$ with respect to the
former. After considering recoils one finds that there is an effective
cutoff on contributions from rapidities (with respect to the thrust
axis) $\eta \gtrsim \ln Q/p_t$, where $p_t$ is the transverse momentum of
the quark with respect to the thrust axis. This leads to $c_B \sim
\langle \ln Q/p_t \rangle$ which is of the order of $ 1/\sqrt{\as}$.

More sophisticated approaches to the problem of power corrections at
first sight seem quite different from the tube model: they examine the
high-order behaviour of the perturbation series, or the dependence of
the observable on a dispersive gluon virtuality (see for example
\cite{DW,DMW,AZ,BB}).  But it turns out that both of these procedures
are equivalent to determining the dependence of the observable on
infra-red properties of the coupling; since the production of gluons
with a given $k_t$ is rapidity-independent and proportional to
$\as(k_t$) we have a situation very similar to the tube model but with
$\phi_h(k_t)$ replaced with $\as(k_t)$.  Accordingly we can write a
relation between $a_0$ and the quantity $\alpha_0(\mu_I)$, often used
in phenomenological analyses, defined as
\begin{equation}
\label{eq:alpha0mom}
\alpha_0(\mu_I) \equiv \int_0^{\mu_I} \frac{dk_t}{\mu_I}\>\as(k_t)\>,
\end{equation}
namely \cite{DW,DMW,Milan,Milan2}
\begin{equation}
  \label{eq:a0Toalpha0}
  \frac{a_0}{Q} = \frac{4C_F}{\pi^2}\cM \frac{\mu_I}{Q}
  \left\{ \alpha_0(\mu_I)- \as(Q) 
    - \beta_0\frac{\as^2}{2\pi}\left(\ln\frac{Q}{\mu_I}+
      \frac{K}{\beta_0}+1\right)\right\}
\end{equation}
with $\cM \simeq 1.490$, $K =
\CA\left(\frac{67}{18}-\frac{\pi^2}{6}\right)-\frac{5}{9}\nf$. The
purpose of the $\as$ terms in \eqref{eq:a0Toalpha0} is to
subtract out contributions that are already taken into account in the
perturbative calculation of the mean value.

It has become a standard procedure to carry out simultaneous fits for
$\as$ and $\alpha_0$ in mean values of a variety of event shapes. One
important test of this class of models for hadronisation corrections
is then that the fitted values $\alpha_0$ (and $\as$) should be the same
for all variables, \ie that $\alpha_0$ should be \emph{universal}.

\subsection{Including mass effects}

From the point of view of the tube model itself nothing changes when
one introduces masses for the hadrons --- we still have a distribution
of hadrons independent of rapidity and with some unknown dependence on
$k_t$. What does change is that we need to use the full (massive)
expressions for the values of the event shapes. In most cases the
event shapes are just defined in terms of the pseudorapidities
(angles) and transverse momenta of the particles. This means that we
have to keep track of the relation between rapidity and
pseudorapidity, which is expressed in the following (equivalent)
equations:
\begin{subequations}
\label{eq:rapVpseudorap}
\begin{align}
  k_t \sinh \etabar &= \sqrt{k_t^2 + m^2} \sinh \eta\,, \\
  k_t^2 \cosh^2 \etabar + m^2 &= (k_t^2 + m^2) \cosh^2 \eta\,.
\end{align}
\end{subequations}
For example for the thrust and the $C$-parameter, we have
\begin{equation}
  \label{eq:NPmassive}
  \langle \cV_{\NP\mathrm{-massive}} \rangle =  \int
  \frac{dk_t}{k_t} \,
  \frac{k_t}{Q} \sum_{h} \phi_h(k_t) \; \int d\eta\,
  f_{\cV}\left(\etabar\left(\eta,\frac{m_h^2}{k_t^2}\right)\right)
\end{equation}
where all that has changed compared to eq.~\eqref{eq:NPmassless} is
the replacement of $f_\cV(\eta)$ with $f_\cV(\etabar)$. However the
fact that $\etabar$ is a function of both $\eta$ and $m^2/k_t^2$ means
that eq.~\eqref{eq:NPmassive} cannot be factorised into two
independent pieces. Hence universality (a direct consequence of the
factorisation) is broken.

For a general variable $\cV$ it will be convenient to write the
resulting non-universal mass-dependent piece of the power correction
as
\begin{equation}
  \label{eq:NPmassiveGen}
  \langle \delta_m \cV \rangle =  \int
  \frac{dk_t}{k_t} \,
  \frac{k_t}{Q} \sum_{h} \phi_h(k_t) \; \int d\eta\;
  \delta f_{\cV}\!\left(\eta,\frac{m_h^2}{k_t^2}\right)\,,
\end{equation}
where for example, for the thrust
\begin{equation}
  \delta f_{\tau}\!\left(\eta,\frac{m^2}{k_t^2}\right) = 
  f_\tau(\etabar) - f_\tau(\eta) = e^{-|\etabar|} - e^{-|\eta|}\,. 
\end{equation}
We then proceed in a manner analogous to that in the massless case. We
define a $\delta c_\cV$,
\begin{equation}
  \label{eq:deltacVmassive}
  \delta c_{\cV}\left(\frac{m^2}{k_t^2}\right) = \int d\eta\, 
  \delta f_{\cV}\!\left(\eta,\frac{m^2}{k_t^2} \right) \,,
\end{equation}
which unlike $c_\cV$ depends on $k_t$ (a consequence of the
non-factorisability). Then the mass-dependent correction to the mean
value of the event shape is
\begin{equation}
  \label{eq:NonUniv}
  \dVmass =  \sum_{h} \frac{m_h}{Q} 
  \int
  \frac{dk_t}{k_t} \,
   \phi_h(k_t)\; \frac{k_t}{m_h}  \,  \delta c_{\cV}
  \left(\frac{m_h^2}{k_t^2}\right) \,.
\end{equation}
To understand the properties of the mass-dependent corrections we need
to study the $\delta c_\cV$'s:
\begin{equation}
\label{eq:deltacVDef}
\delta c_\cV\left(\frac{m^2}{k_t^2}\right)
 =  \left\{ \begin{array}[c]{ll}
  \displaystyle\int d\etabar \,f_{\cV}(\etabar)\, \left(
    \frac{d\eta}{d\etabar} - 
    1\right) & \cV = \tau,C,B_T,B_W,\vspace{0.2cm}\\
  \displaystyle\int d\eta \,e^{-\eta}\,\Theta(\eta)  \left(\sqrt{1 +
      m^2/k_t^2} - 1\right) & 
  \cV = \rho,\rho_h\,,
  \end{array} \right.
\end{equation}
where
\begin{equation}\label{eq:detadetabar}
  \frac{d\eta}{d\etabar} = \frac{\cosh \etabar}{\sqrt{\cosh^2 \etabar
      + m^2/k_t^2}}\,.
\end{equation}
For the event-shapes in the first line of eq.~\eqref{eq:deltacVDef}
the integrals have been rewritten with a change of variable in the
first term as this simplifies their subsequent evaluation.

The exact forms for the $\delta c_\cV$ are
\begin{subequations}
\begin{align}
  \delta c_\tau \left(\frac{m^2}{k_t^2}\right) &= 2\left[
    \frac{1}{\xi} \,\bK\! \left(\frac{\sqrt{\xi^2-1}}{\xi}\right) -
    \xi \bE\!\left(\frac{\sqrt{\xi^2-1}}{\xi}\right) + \xi
    - 1 \right],\\
  \delta c_C \left(\frac{m^2}{k_t^2}\right) &= \frac{6}{\xi}
  \,\bK\!\left(\frac{\sqrt{\xi^2-1}}{\xi}\right) - 3\pi
  \,,\\
  \delta c_{B_T} \left(\frac{m^2}{k_t^2}\right) &= 2\delta c_{B_W} =
  -\ln \xi\,,\\
  \delta c_\rho\left(\frac{m^2}{k_t^2}\right) &= \delta c_{\rho_h} =
  \xi - 1\,,
\end{align}
\end{subequations}
where we have introduced the shorthand $\xi^2 = 1 + m^2/k_t^2$, and
$\bE$ and $\bK$ are the complete elliptic integrals defined as follows,
\begin{subequations}
\begin{align}
  \bE(x) &= \int_0^{\pi/2} \sqrt{1 - x^2 \sin^2 \psi}\; d\psi \qquad
  (|x| \le 1)\,, \\
  \bK(x) &= \int_0^{\pi/2} \frac{d\psi}{\sqrt{1 - x^2 \sin^2
      \psi}}\qquad \quad \;(|x| <1)\,.
\end{align}
\end{subequations}

\begin{figure}[tb]
  \begin{center}
    \epsfig{file=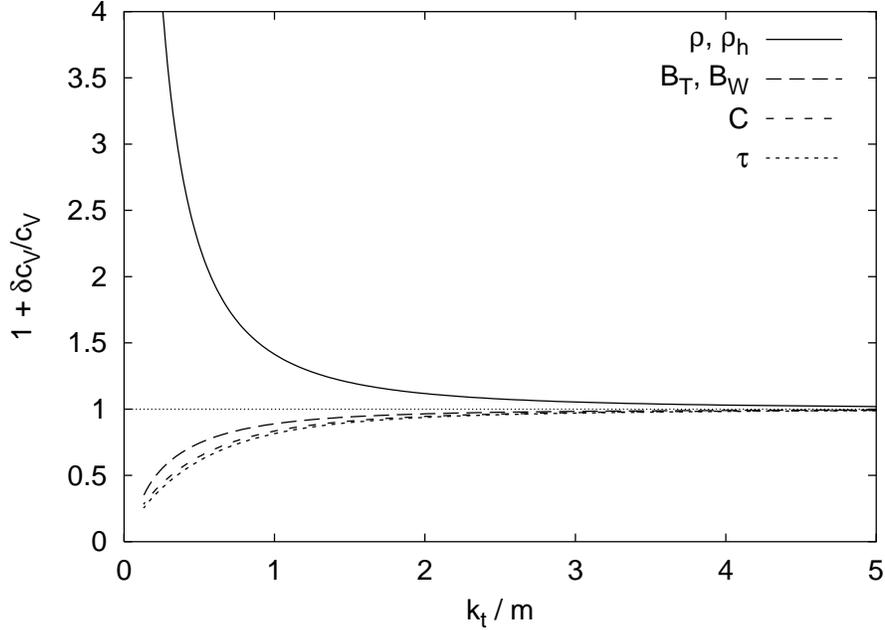,width=0.7\textwidth}
    \caption{The dependence of $c_\cV + \delta c_\cV$ on particle
      masses, shown for the four distinct groups of variables.}
    \label{fig:anlmass2}
  \end{center}
\end{figure}

To see how the $\delta c_\cV$ will affect our power correction we
compare them to the universal $c_\cV$ power contribution.
Figure~\ref{fig:anlmass2} shows $[c_\cV + \delta c_\cV(m^2/k_t^2)] /
c_\cV$ for a range of variables as a function of $k_t/m$. One
immediately sees that for small $k_t$ the non-perturbative correction
to the jet mass will be enhanced, while the NP correction to the other
variables will be suppressed. Furthermore the enhancement for the jet
mass is much larger than the suppression for the other variables
(which are fairly similar to one another).

\begin{table}[tb]
\begin{center}
  \begin{tabular}{|c||c|c|c|c|c|c|}
    \hline
$\cV$ & $\tau$ & $\rho$ & $\rho_h$ & $C$  & $B_T$ & $B_W$    
     \\  \hline
$\gamma_\cV$ &
     $-\pi/2$+1& $1/2$ & $1/2$ & $-3\pi/4$ & $-1/2$ & $-1/4$ 
     \\ \hline
$\gamma_\cV/c_\cV$ &
     $-0.285$ & $0.5$ & $0.5$ & $-0.25$ & $-0.180$ & $-0.227$
    \\ \hline
  \end{tabular}
\end{center}
\caption{The coefficients of the high-$k_t$ behaviour of the
  mass-dependent power correction, 
  $\gamma_\cV$. In the case of the broadenings the $c_\cV$ depend on
  $\as(Q)$, so the ratio $\gamma_\cV/c_\cV$ is shown for $Q = M_Z$.} 
\label{tab:gammas}
\end{table}
To study the question more quantitatively we observe that for large
$k_t$, the $\delta c_\cV$ scale as $m^2/k_t^2$,
\begin{equation}
  \delta c_\cV\left(\frac{m^2}{k_t^2}\right)
  = \gamma_\cV \frac{m^2}{k_t^2} + \cO{\frac{m^4}{k_t^4}}
  \label{eq:GammaExpnd}
\end{equation}
with the $\gamma_\cV$ given in table~\ref{tab:gammas}.  This means
that the integral in eq.~\eqref{eq:NonUniv} is dominated by low
momenta for all reasonable forms of the distribution of particles
$\phi_h(k_t)$, and hence just gives a number. Therefore $\dVmass$ is
proportional to $1/Q$, \ie formally of the same order as the universal
power correction.

\begin{figure}[tbp]
  \begin{center}
    \epsfig{file=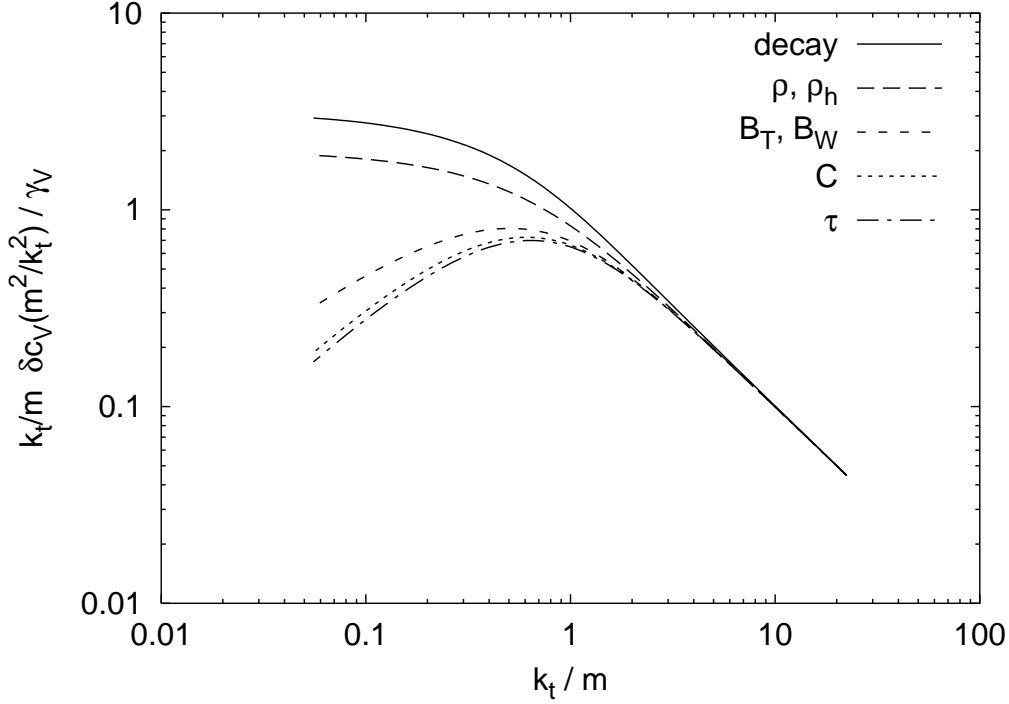,width=0.8\textwidth}
    \caption{The observable-dependent piece of the integrand for the
      non-universal mass-dependent power correction,
      eq.~\eqref{eq:NonUniv}, normalised to $\gamma_\cV$. The curve
      labelled `decay' will be discussed in
      section~\ref{sec:decayscheme}.}
    \label{fig:DeltaCV}
  \end{center}
\end{figure}

The question of the quantitative relationship between the sizes
of the mass-corrections in the different observables is more delicate
because it depends on the region of $k_t$ which dominates the integral
in eq.~\eqref{eq:NonUniv}. If $\phi(k_t)$ is such that moderate
$k_t$'s dominate (\ie where $\delta c_\cV$ is equal to $\gamma_\cV
m^2/k_t^2$) then we can expect the following relation to hold:
\begin{equation}
  \langle \delta_m \cV \rangle \propto \gamma_\cV\,.
\end{equation}
If $\phi(k_t)$ is such that smaller $k_t$'s dominate the integral
\eqref{eq:NonUniv}, then formally we can make no such statement.
Nevertheless by examining figure~\ref{fig:DeltaCV}, which shows
$k_t/m \delta c_\cV/\gamma_\cV$ as a function of $k_t/m$, one observes that
for $\tau$, $C$ and the broadenings the shape of the $\delta
c_\cV(m^2/k_t^2)$ functions are very similar. This means that
regardless of the form of $\phi(k_t)$ we will still observe the
property that
\begin{equation}
  \label{eq:ratioNonUniv}
  \frac{\langle \delta_m \cV_{1} \rangle}{\gamma_{\cV_1}}
   \simeq \frac{\langle \delta_m \cV_{2}
    \rangle}{\gamma_{\cV_2}}\,.
\end{equation}
For the jet masses on the other hand, $\delta c_\cV(m^2/k_t^2)$ has a
different shape and being considerably larger at small momenta, so
that we can expect the following relation to hold:
\begin{equation}
  \label{eq:ratioNonUnivRho}
  \frac{\langle \delta_m \rho \rangle}{\gamma_{\rho}}
   \gtrsim \frac{\langle \delta_m \tau \rangle}{\gamma_{\tau}}\,.
\end{equation}

The conclusion of this section is that mass effects introduce extra
$1/Q$ power corrections, which break the simple `universal' picture of
power corrections that is obtained in the massless case. For most
variables we expect a negative correction, whose magnitude is roughly
proportional to $\gamma_\cV$ (which for these variables is roughly
$-c_\cV/4$). For the jet masses we expect positive corrections whose
magnitude is larger than what would be expected from a simple
proportionality to $\gamma_\cV$ (which itself is $c_\cV/2$) .

\subsection{Alternative schemes}

So far we have used the event-shape definitions given in
eqs.~\eqref{eq:ShapeDefs}. From the point of view of the perturbative
QCD calculation however we could have chosen any number of related
definitions with the same massless limit and we would have obtained
the same perturbative (and universal non-perturbative) predictions.
Here we discuss two particular examples of such modifications.

\paragraph{The $\boldsymbol{p}$-scheme:} The difference between the jet
masses and the other variables occurs because the jet masses are the
only variables to be sensitive to the difference between the energies
and 3-momenta of the particles.  However one could equally well
consider a second pair of variables, identical to the jet masses
except that they are defined only in terms of the particle 3-momenta
(\ie in the definition, each occurrence of particle energy is replaced
by the modulus of the corresponding 3-momentum). We will refer to these
as the jet masses in the $p$-scheme (whereas we will refer to the
default definitions as the massive scheme).

As pointed out at the beginning of the section, from the point of view
of the perturbative and universal non-perturbative calculations, which
ignore particle masses, such a variable would have identical
properties to the original jet mass.  However, for an event consisting
of soft massive particles its value would be
\begin{equation}
    \rho_{p} \simeq \frac{1}{Q} \sum_{i \in
      \mathrm{soft}} k_{ti}\, e^{-|\etabar_i|} \,\Theta(\etabar_i)\,,
\end{equation}
rather than eq.~\eqref{eq:rhoSoft}. Noting the similarity between this 
expression and eq.~\eqref{eq:tauSoft}, one obtains that
\begin{equation}
  \frac{\delta c_{\rho_{p}}(m^2/k_t^2)}{c_\rho}
    = \frac{\delta c_{\tau} (m^2/k_t^2)}{c_\tau}\,,
\end{equation}
\ie relative to the universal power correction the mass-dependent
piece is identical in the two cases.  The use of the $p$-scheme makes
no difference for the variables other than jet masses, since they are
all already defined purely in terms of the 3-momenta.

So in the $p$-scheme \emph{all} variables should have a mass-dependent
correction which is roughly proportional to $\gamma_\cV$, which itself
is roughly proportional to $c_\cV$. Therefore universality should (more or
less) appear to remain intact.

\paragraph{The $\boldsymbol{E}$-scheme:} Another definition of
event-shapes which is identical at the perturbative and universal
non-perturbative level is one defined purely in terms of the energies
and directions of particles, \ie where all 3-momenta are substituted
with momenta in the same direction but whose modulus is equal to the
energy. We call this the $E$-scheme. We note that similar definitions,
in terms of energy flow, have been suggested in the past by various
authors \cite{KS,Tkachov}, on the grounds that they are closer to what
is measured in a experimental calorimeter and that they may also allow
event shapes to be expressed in terms of correlation function of
fields.

The expression for $\rho_{E}$ in the soft limit is
\begin{equation}
    \rho_{E} \simeq \frac{1}{Q} \sum_{i \in
      \mathrm{soft}} k_{ti}\, e^{-|\etabar_i|} \,\Theta(\etabar_i)
     \, \frac{\sqrt{\cosh^2 \etabar_i + m^2_i/k_{ti}^2}}{\cosh \etabar_i}\,,
\end{equation}
where the extra factor compared to $\rho_{p}$ is the
ratio of the energy to the 3-momentum. The expression for $\delta
c_{\rho_{E}}$ is then
\begin{equation}
  \delta c_{\rho_{E}}\left(\frac{m^2}{k_t^2}\right)
   = \int d\etabar \, e^{-|\etabar|} \,\Theta(\etabar) \left(
    \frac{d\eta}{d\etabar} \cdot \frac{\sqrt{\cosh^2 \etabar +
        m^2/k_{t}^2}}{\cosh \etabar}  - 1\right)\,.
\end{equation}
However, noting the form of $d\eta/d\etabar$,
eq.~\eqref{eq:detadetabar}, one sees that this is identically zero. A
similar phenomenon occurs for the other variables, \ie in general we
have
\begin{equation}
  \delta c_{\cV_{E}}\left(\frac{m^2}{k_t^2}\right)
  \equiv 0\,.
\end{equation}
In other words in the $E$-scheme there is no non-universal
mass-dependent $1/Q$ power correction. So if one wants to study the
universality of $1/Q$ power corrections, the best to way to do it is
to measure all variables in the $E$-scheme.\footnote{We note though one
  small defect of the $E$-scheme, namely that the rescaled 3-momenta
  do not necessarily add up to exactly zero. Experimentally this is in
  any case quite common due to measurement errors, and so is not
  necessarily a major defect. However to preserve various desirable
  properties of the event-shape definitions, in the $E$-scheme we
  choose to boost the event (by a small amount, of order $\Lambda/Q$)
  so as to place it in the centre-of-mass frame. One might worry that
  the boost itself might alter the value of the event shape, but it
  can be shown that the effect of the boost is to modify the event
  shape by a \emph{relative} amount $\Lambda/Q$, so that the effect on
  the mean value is of order $\as(Q) \Lambda/Q$, \ie formally
  negligible.}

The $p$ and $E$-schemes are of also of interest because in principle
one can measure the difference between a given observable in two
different schemes. For example if one measures the difference between
$\tau$ in the $p$ and $E$-schemes, one expects this to be equal to
$\langle\delta_m \tau\rangle$. This can be done for various observables, after which
one can verify the relations
(\ref{eq:ratioNonUniv}, \ref{eq:ratioNonUnivRho}).

\subsection{Hadron decay}
\label{sec:decayscheme}

Quite often hadron-level measurements are performed on particles which
are unstable, though long-lived compared to their time of flight
across the detector. The definition of the observable does not specify
at what stage of the hadron decay chain we should make our
measurement, so we are actually free to make it at any stage we like
--- as long as we specify the stage.

This leads us to wonder about dependence of the observable on the
particular hadron level that is chosen. It is possible quite generally
to argue that redefining the hadron level should not affect the
universality pattern. Suppose one starts off at a stage consisting of
short-lived hadronic resonances. The boost-invariant nature of the
mechanisms of hadronisation implies that these hadronic resonances
will have been produced with a rapidity-independent distribution.
Hadron decay is also a boost invariant process, so the decay products
of the resonances will also be distributed in a rapidity-invariant
manner. But the mean transverse momentum may well change because the
decay of a hadron liberates energy, some of which may enter into the
transverse degrees of freedom. Therefore the power correction to a
given observable will increase as a result of hadron decay, but the
increase will be proportional to $c_\cV$ (as long as the observable
is measured in the $E$-scheme).

Though it is perhaps more natural to treat hadron decay as a change in
the distributions $\phi_h(k_t)$, if one wants quantitative predictions
it turns out to be more convenient to discuss it in terms of a $\delta
c_\cV(m^2/k_t^2)$ contribution. Since our decay process is rapidity
independent we can write the $\delta c_{\cV_{\decay}}$ for an arbitrary
variable as
\begin{equation}
  \label{eq:deltaCdecay}
  \delta c_{\cV_{\decay}}
  \left(\frac{m^2}{k_t^2}\right) =  c_\cV\, X_{\decay}
  \left(\frac{m^2}{k_t^2}\right) 
\end{equation}
where $X_{\decay}$, which is independent of the variable, is defined by
\begin{equation}
  \label{eq:genXdef}
  X_{\decay} \left(\frac{m^2}{k_t^2}\right) = 
  \left\langle \frac{1}{k_{t}} \sum_{i\,\in\,\mathrm{children}}
    |k_{ti}|\right\rangle - 1\,,
\end{equation}
is the mean relative change in total transverse momentum as a result
of the decay of a parent with transverse momentum $k_t$ and mass $m$.

In general $X_\decay$ is a fairly complicated function, because
actual hadronic decays can involve 3-body final states where one or
more of the decay products is massive. For the purposes of the studies
in this paper we instead introduce a decay-scheme, in which all
hadrons are \emph{artificially} decayed to a pair of massless
particles. At first sight this seems a little arbitrary, but there are
two reasons why it is nevertheless of interest. Firstly, it is fairly
straightforward to apply this scheme to a given ensemble of particles:
one simply takes each hadron and in its centre of mass frame decays it
to two particles moving in opposite directions along a randomly chosen
axis. The second reason is phenomenological: as we shall discover in
section~\ref{sec:reshaddec} if one applies the procedure to different
hadron levels (say the normal hadron level and some `resonance' level,
earlier on in the decay chain) one finds that the decay scheme results
in the two cases are very similar --- in other words, decay-scheme
results are almost independent of the particular hadron level from
which one starts.

In this 2-body decay scheme, $X_\decay$ is given by the following
expression 
\begin{multline}
  \label{eq:Xdecay}
  X_{\decay}
  \left(\frac{m^2}{k_t^2}\right) = \\ \qquad
   \int_{-1}^{1} d(\cos\theta) \int_0^{2\pi} \frac{d\phi}{4\pi} 
   \left(
     \sqrt{1 + \left(\frac{m^2}{k_t^2}+ \cos^2\phi\right)\sin^2\theta
       + 2\sqrt{1 + \frac{m^2}{k_t^2}}
       \sin\theta \cos\phi}  
      - 1\right),
\end{multline}
for which we have yet to find a closed form. Its behaviour for $k_t\gg
m$ is such that for a general variable the expansion coefficient
defined in eq.~\eqref{eq:GammaExpnd} is
\begin{equation}
  \label{eq:gammaVdecay}
  \gamma_{\cV_{\mathrm{decay}}} = \frac14  c_\cV\,.
\end{equation}
The shape of the function $\delta c_{\cV_{\mathrm{decay}}}$ is similar
to that of $\delta c_\rho$, as can be seen from
figure~\ref{fig:DeltaCV}.

\subsection{Relation to massive-gluon calculations}
\label{sec:massivegluoncalc}

Many of the traditional power-correction calculations are based on the
dispersive, or massive-gluon approach. Often however, as we have
already mentioned, there is an implicit or even explicit
\cite{\allmilan} assumption that the gluon decays into massless
particles, leading to the statement of universality.

One exception is the calculation in \cite{BB} which considered the
thrust using the full kinematics of the undecayed massive gluon (both
in the numerator, where gluon decay often makes no difference, and in
the denominator where it does make a difference). They obtained the
result that the coefficient $c_\tau$ for the thrust should be $4 G$
rather than $2$, where $G\simeq 0.916$ is Catalan's constant. This
seems quite strange since we have argued that a proper treatment of
particle masses should \emph{lower} the value of the thrust rather
than increase it, whereas the analysis of \cite{BB} suggests that the
power correction increases. However the situation is subtle because
the massive-gluon approach takes the power correction as being
proportional to the non-analyticity in the gluon mass (or virtuality)
after integration over the whole of phase space, and there is a
non-trivial relation between the effect of the gluon mass on the value
of $\tau$ for a given event and the non-analyticity.

\section{QCD-based analysis}
\label{sec:QCD}

When discussing universal power corrections in section~\ref{sec:tube}
we introduced the function $\phi_h(k_t)$, the one-particle inclusive
distribution for the non-perturbative production of hadrons $h$ with
transverse momentum $k_t$. Within the tube model we vaguely know what
we mean (the hadronisation associated with the low momentum fields
from the $q,\qbar$ sources), but in QCD it is quite ambiguous: after
all, all hadrons are produced non-perturbatively!

Strictly what we are interested in is the difference between our
observable at the hadron level and the value calculated from a given
order of perturbation theory. In our particular case the variables are
sensitive to the mean transverse momentum (at a given rapidity). A
proper definition of the difference in mean transverse momentum
between parton and hadron levels is
\begin{equation}
  \label{eq:a0RigDef}
  a_0 =  \int \frac{d k_t}{k_t} k_t \left(
    \sum_h \Phi_h(k_t,\eta) - \sum_{p=q,g} \Phi_p(k_t,\eta) \right)
\end{equation}
where $\Phi_h(k_t,\eta)$ is the distribution of \emph{hadrons} $h$ at
a given transverse momentum and rapidity, and $\Phi_{q,g}(k_t,\eta)$
is the perturbative distribution of \emph{partons}. So the quantity
$\phi_h(k_t)$ should really be understood as being defined as follows:
\begin{equation}
  \sum_{h} \phi_h(k_t) \cong \sum_h \Phi_h(k_t,\eta) - \sum_{p=q,g}
  \Phi_p(k_t,\eta).
\end{equation}

In eq.~\ref{eq:a0RigDef} the integrals over $\Phi_h(k_t,\eta)$ or
$\Phi_{q,g}(k_t,\eta)$ separately would have values of the order of
$\as\, e^{-|\eta|} \,Q$, because a fraction $\as$ of the time there can
be a hard particle. But typical pictures of hadronisation state that
the difference should be dominated by particles `produced at low
transverse momenta,' as opposed to particles which come from the
collinear fragmentation of a hard parton, since in the latter case the
sum of $k_t$'s of the hadrons should on average be equal to the $k_t$
of the original parton and there will be no contribution to the
difference \eqref{eq:a0RigDef}.  This ensures that the integral of
$\sum_h \phi_h$ is dominated by low $k_t$'s and that it is roughly
rapidity and $Q$-independent.

But when working out mass-dependent effects, $\Phi_{q,g}$ does not
contribute at all since in the perturbative calculation one has
massless quarks and gluons (we do not consider the case of
calculations with massive quarks). So in the expression for the
mass-dependent non-universal power correction,
eq.~\eqref{eq:NPmassive}, we shall replace $\phi_h(k_t)$ with
$\Phi_h(k_t,\eta)$.

\subsection{Spectrum of hadrons}
\label{sec:spect}
To study mass effects in detail it is necessary to have some
understanding about $\Phi_h$. The simplest approach that is currently
available is based on local parton-hadron duality (LPHD)
\cite{DTProc,LPHD}, 
namely the idea that on average there is a correspondence between the
production of partons and the production of hadrons. One can then
calculate the distribution of partons and expect the distribution of
hadrons to be very similar. Using this idea the distribution of
low-$k_t$ hadrons has been calculated as a function of $k_t$ and
rapidity in \cite{KLO}:
\begin{equation}
  \label{eq:PhiKLO}
  \Phi(k_t, \eta, \ln Q/\Lambda) \sim \frac{4\CF}{\beta_0} \left(
      \frac1{\ln k_t/\Lambda}
    + \frac{4\CA}{\beta_0} \ln \frac{\ln k_t/\Lambda}{\ln Q_0/\Lambda}
              \ln \frac{\ln Q e^{-|\eta|}/\Lambda}{\ln k_t/\Lambda}
    + \cdots \right)
\end{equation}
where $\Lambda$ is the QCD scale in some arbitrary scheme, $Q_0$ is an
unknown cutoff below which parton branching stops. We have explicitly
added $\ln Q$ as an argument of $\Phi$ to emphasise that it is now $Q$
dependent. The $\eta$ dependence of $\Phi$ is actually properly
described by this formula only for large $\eta$. But since most of our
integrals in $\eta$ converge rapidly we will usually be able to ignore
the $\eta$-dependence altogether.

The first term in the brackets in \eqref{eq:PhiKLO} just corresponds
to the radiation of a single gluon of transverse momentum $k_t$ from
the $q\qbar$ pair, with intensity $\as(k_t)$.  This term is both
rapidity and $Q$-independent.  The second term comes from the coherent
(or angular ordered) radiation of another gluon from the first gluon,
with the logarithms originating from the integrations over the two
gluon momenta. 

At lowest momenta the first term dominates. At higher momenta the
second term becomes more important. We note that it has significant
$Q$-dependence: this means that a piece of the mass-dependent
correction will behave as $\ln \ln Q$ times some function of $k_t$,
and the fact that this function is enhanced at larger values of $k_t$
implies sensitivity to $\delta c_\cV(m^2/k_t^2)$ in a
region where the approximation of $\delta c_\cV$ by $\gamma_\cV\,
m^2/k_t^2$ might be expected to work --- in other words we expect
there to be a term in the mass-dependent power correction proportional
to
\begin{equation*}
  \gamma_\cV \, \frac{\ln \ln Q/\Lambda}{Q}\,,
\end{equation*}
as opposed to a simple $1/Q$ correction. 

A proper treatment requires that one take into consideration not only
the first term to have $Q$ dependence in \eqref{eq:PhiKLO} but also
yet higher orders.  This can be done via moments $D_\om(\ln Q)$ of the
multiplicity distribution of particles with momentum fraction $x$
emitted from a gluon at scale $Q$, $D(x,\ln Q)$:
\begin{equation}
  \label{eq:MomMult}
  D_\om(Y=\ln Q/\Lambda) = \int_0^1 \frac{dx}{x} x^\om \left[ x D(x,Y)
  \right]\,. 
\end{equation}
The corresponding moment for emissions from a quark is just $\CF/\CA
D_\om(Y)$. We will actually be interested in the moments of the
multiplicity distribution at fixed $\eta$ (which we take positive),
$\Phi_\om(\eta)$,
\begin{equation}
  \Phi_\om(\eta, \ln Q/\Lambda) = \int \frac{dk_t}{k_t}
  \left(\frac{k_t}{Q}\right)^\om \Phi(k_t,\eta, \ln Q/\Lambda) 
\end{equation}
which is given in terms of $D_\om$ by
\begin{equation}
  \Phi_\om(\eta, \ln Q/\Lambda) = e^{-\om \eta}
  \left. \frac{d}{dY} D_\om(Y) \right|_{Y =
    \ln Q/\Lambda - \eta}\,.
\end{equation}
The full multiplicity moment satisfies the following equation,
embodying coherence \cite{coherence} at double logarithmic accuracy
(DLA) as discussed for example in \cite{Basics}
\begin{equation}
  \label{eq:diffDom}
  \frac{d}{dY} D_\om(Y) = \int_0^\infty dy \,e^{-\om y}\, 4\CA
  \frac{\as(Y-y)}{2\pi} D_\om(Y-y)\,.
\end{equation}
By differentiating both sides this can be written as a second order
differential equation, for which an approximate (DLA) solution is
\begin{equation}
  \label{eq:Dom}
  D_\om(Y) \simeq D_\om(Y_0) \exp\left[ \int_{Y_0}^Y dy\,
    \gamma_\om^{DLA}(\as(y))\right]\,, 
\end{equation}
with
\begin{equation}
  \label{eq:gammaom}
  \gamma_\om^{DLA}(\as) = \frac12\left( - \om + \sqrt{\om^2 +
      \frac{16\CA \as}{2\pi}}\right)\,.
\end{equation}
Hence the moment of the distribution at fixed rapidity is given by
\begin{equation}
  \Phi_\om(\eta,\ln Q/\Lambda) = \gamma_\om D_\om(\ln Q/\Lambda -
  \eta) e^{-\omega \eta}   \,.
\end{equation}
Eq.~\eqref{eq:diffDom} and its solution eq.~\eqref{eq:gammaom} are
usually derived for the region around $\om=0$. There 
they are known to give the correct leading term of $\gamma_\om$,
proportional to $\sqrt{\as}$. The first set of subleading corrections in
this region are also known and can be obtained within the modified
leading log approximation (MLLA) which takes into account effects such
as a gluon splitting into quarks, and the part of the $P_{gg}$
splitting function which is finite at $z\to0$. These corrections can
be embodied into a modification of $\gamma_\om$ and give
\cite{DTProc,MLLA}
\begin{equation}
  \label{eq:gammaMLLA}
  \gamma_\om(\as) = \gamma_\om^{DLA}(\as) + 
  \frac{\as}{2\pi}\left[-\frac{a}{2}\left(1 +
      \frac{\om}{\sqrt{\om^2+16\CA\as/2\pi}}\right) + \beta_0
    \frac{16\CA\as/2\pi}{\om^2 + 16\CA\as/2\pi}
  \right] \,,
\end{equation}
where $a = 11\CA/3 + 2\nf/3\CA^2$. Around $\om=0$ MLLA effects give
corrections to $\gamma_\om$ of order $\as$, \ie suppressed by an
amount $\sqrt{\as}$ compared to the leading contribution; effects
associated with the correct scale choice for $\as$ start at
${\cal O}(\as^{3/2})$ and so do not mix with the MLLA corrections.

For our applications we are actually interested in the region around
$\om=-1$ and it is not immediately obvious that we can apply the
$\om\sim0$ derivation.  One can envisage two sources of problems:
firstly since the $D_{\om=-1}$ moment is dominated by low momenta one
might worry that its evolution is entirely non-perturbative. Secondly
one may wonder whether the soft approximation of the $P_{gg}$
splitting function, implicitly included in eq.~\eqref{eq:diffDom}, is
valid (for $\om=1$ for example it would not be). But bearing in mind
that for negative $\om$, $\gamma_\om(\as) = -\om + \order{\as}$, we
see that eq.~\eqref{eq:diffDom} has an integrand nearly independent of
$y$ over the whole integration region (modulo powers of $\as(Y-y)$),
so that the integration is logarithmic. This means that it is
dominated neither by the very soft (non-perturbative) region, nor by
the region in which the splitting is hard, and as a consequence it is
safe to write eq.~\eqref{eq:diffDom}. However since we are not in a
double logarithmic, but a single logarithmic region, we can only trust
the first order expansion of $\ga_\om$:
\begin{equation}
  \label{eq:gammaomNonZero}
  \gamma_\om(\as) =  -\om - \frac{4\CA \as}{2\pi\om} + \order{\as^2},
  \quad\qquad {\om < 0}\,.
\end{equation}
Pieces of order $\as^2$ come additionally both from the MLLA
corrections and from other uncalculated sources such as the scale
choice for $\as$, which is beyond our control.  This means that we are
not able to go beyond leading order in our studies.

Now that we have an expression for $\gamma_\om(\as)$, we make the
standard assumption that branching only occurs above some scale $Q_0$,
so that $D_\om(Y_0 = \ln Q_0/\Lambda) = 1$ and we arrive at the results
\begin{equation}
  \label{eq:Domsim}
  D_{\om}(Y) \sim e^{-\om(Y-Y_0)}
  \left(\frac{Y}{Y_0}\right)^{-\frac{A}{\om} }\,, 
\end{equation}
and
\begin{equation}
  \label{eq:Phiom}
  \Phi_{\om}(\eta,Y) \sim -\om e^{-\om(Y-Y_0)}
  \left(\frac{Y-\eta}{Y_0}\right)^{-\frac{A}{\om} }\,,
\end{equation}
where $A = 4\CA/\beta_0 \simeq 1.565 (1.714)$ for $\nf = 5(6)$.

The $D_\om$ and $\Phi_\om$ obtained so far correspond to expectations
for numbers of gluons. The LPHD hypothesis suggests that for a given
hadron species $h$ we should have,
\begin{equation}
  \Phi_{h,\om}(\eta,Y) = - \cN_{h,\om} \, \om\, e^{-\om(Y-Y_{0,h})}
  \left(\frac{Y - \eta}{Y_{0,h}}\right)^{-\frac{A}{\om} }\,.
\end{equation}
where $\cN_{h,\om}$ is a unknown normalisation factor, which contains
the information about the conversion of partons into a given hadron
species $h$. It should depend on $Y_{0,h} = \ln Q_{0,h}/\Lambda$ in
such a way as to ensure that the final result for $\Phi_{h,\om}$ is
independent of $Q_{0,h}$. The assumption of local parton-hadron duality
implies that $\cN_{h,\om}$ is free of soft divergences, since these
should all have been taken into account in the QCD treatment of gluon
radiation.

\subsection{Application to power corrections}
\label{sec:QCDappl}

The expression for the mass-dependent piece of the power correction is
(\cf eq.~\eqref{eq:NPmassiveGen})
\begin{equation}
  \label{eq:fullDeltaMV}
  \dmass{\cV} =  \sum_{h} 
  \int
  \frac{dk_t}{k_t} \,
    \frac{k_t}{Q} \int d\eta \;
    \delta f_\cV\!\left(\eta, \frac{m_h^2}{k_t^2}\right)\, 
    \Phi_h(k_t,\eta, Y)\,,
\end{equation}
which we can rewrite as
\begin{equation}
  \label{eq:NonUnivOmegaRap}
  \dmass{\cV} =  \sum_{h} 
  \int
  \frac{d\om}{2\pi i} \, \int d\eta\,
    \,\delta f_{\cV,\om} (\eta)\,
    \Phi_{h,-\om}(\eta,Y)\,,
\end{equation}
where we have defined (note the extra factor of $k_t/Q$)
\begin{equation}
  \delta f_{\cV,\om}(\eta) =  \int \frac{dk_t}{k_t}
  \left(\frac{k_t}{Q}\right)^\om \;\frac{k_t}{Q} \,
  \delta f_\cV\!\left(\eta, \frac{m_h^2}{k_t^2}\right)\,.
\end{equation}
Let us then expand the rapidity dependence of $
\Phi_{h,-\om}(\eta,Y)$:
\begin{equation}
  \Phi_{h,-\om}(\eta,Y) = \Phi_{h,-\om}(0,Y) \left(1 -
    \frac{A}{\om}\frac{\eta}{Y}  + \cdots \right)\,.
\end{equation}
We see that rapidity dependent pieces are suppressed by powers of
$1/Y$, or equivalently by powers of $\as$. Since for most variables
(the special case of the broadenings is discussed in
appendix~\ref{sec:broad}) the rapidity integration in
eq.~\eqref{eq:NonUnivOmegaRap} converges rapidly, powers of $\eta$ do
not lead to any particular enhancement and we can simply neglect the
rapidity dependence of $\Phi_{h,-\om}$:
\begin{equation}
  \label{eq:NonUnivOmega}
  \dmass{\cV} =  \sum_{h} 
  \int
  \frac{d\om}{2\pi i} \,
    \,\delta c_{\cV,\om}\,
    \Phi_{h,-\om}(\eta=0,Y)\,,
\end{equation}
where we have defined
\begin{equation}
  \delta c_{\cV,\om} =  
  \int d\eta \, \delta f_{\cV,\om}(\eta) = 
  \int \frac{dk_t}{k_t}
  \left(\frac{k_t}{Q}\right)^\om \;\frac{k_t}{Q} \,
  \delta c_\cV\!\left(\frac{m_h^2}{k_t^2}\right)\,.
\end{equation}
To understand the structure of eq.~\eqref{eq:NonUnivOmega} let us
consider for now just the case of the jet mass (in its default,
massive scheme), which has
\begin{equation}
  \delta c_{\rho,\om} = - \left(\frac{m_h}{Q}\right)^{\om+1} 
         \frac{\Gamma\left(\frac{\om}{2}\right)
                    \Gamma\left(-\frac{1+\om}{2}\right)
                    }{4\sqrt{\pi}}\,.
\end{equation}
Eq.~\eqref{eq:NonUnivOmega} then becomes
\begin{equation}
  \label{eq:NonUnivOmegaRho}
  \dmass{\rho} =  \sum_{h}  \frac{m_h}{Q}
  \int
  \frac{d\om}{2\pi i} \,\cN_{h,-\om} \,\om\,
    \frac{\Gamma\left(\frac{\om}{2}\right)
           \Gamma\left(-\frac{1+\om}{2}\right)
            }{4\sqrt{\pi}} 
    \left(\frac{m_h}{Q_{0,h}} \right)^{\om}
          \left(\frac{\ln Q/\Lambda}{\ln Q_{0,h}/\Lambda}
            \right)^{\frac{A}{\om}}\,,
\end{equation}
where the integration contour passes between $\om=0$ and $\om=1$. For
sufficiently large $Q$ the integrand has a saddle point close to
$\om=1$ and accordingly we consider its behaviour in that
region:\footnote{assuming $\cN_{h,-\om}$ to be free of
  non-analyticity around $\om=1$.}
\begin{equation}
  \frac{\gamma_\rho\, \cN_{h,-1}}{1-\om}\,
  \left(\frac{m_h}{Q_{0,h}} \right)^{\om}
  \left(\frac{\ln Q/\Lambda}{\ln Q_{0,h}/\Lambda}
  \right)^{A(2-\om)}.
\end{equation}
Such a form holds in general for all the variables.  If we use it to
evaluate the saddle-point integral we obtain the following result 
\begin{equation}
  \label{eq:dcvAnswer}
  \dmass{\cV} \simeq \gamma_\cV \sum_{h} \cN_{h,-1} \frac{m_h^2}{Q Q_{0,h}} 
       \left(\frac{\ln Q/\Lambda}{\ln Q_{0,h}/\Lambda}
  \right)^{A}
\end{equation}
When $m_h<Q_{0,h}$, we can quite easily study the corrections to this
result since the contour in \eqref{eq:NonUnivOmegaRho} can be closed
to the right, and the integral is equal to the sum of residues at
$\om=1,3,5,\ldots$.  The first residue just gives our answer
\eqref{eq:dcvAnswer}. The relative magnitude of the contribution from
higher residues depends critically on the normalisation of the poles
of $\delta c_{\cV,\om}$ at $\om=3,5,\ldots$ (which can be worked out)
and on the value of $m_h/Q_{0,h}$ (which is unknown) and so cannot be
determined a priori.  However the energy dependence of these higher
residues is much weaker: for example the residue at $\om=3$ goes as
$(\ln Q/\Lambda)^{A/3}$. We are of course assuming that $\cN_{h,-\om}$
has no relevant non-analytic structure. This cannot be guaranteed, and
for example if the distribution of hadrons goes as
$dk_t^2/(k_t^2+m^2)$ then we expect $\cN_{h,-\om}$ to have a pole at
$\om=2$. This would lead to corrections to our results proportional to
$(\ln Q/\Lambda)^{A/2}$.

So we expect that the mass-dependent power correction should go as
\begin{equation}
  \label{eq:Final}
  \dmass{\cV} = \frac{\gamma_\cV \,\epsilon}{Q} \,\ln^A
  \frac{Q}{\Lambda}
    + \order{
    \frac{\Lambda}{Q} \, \ln^{A/3} \frac{Q}{\Lambda} } 
\end{equation}
where $A/3$ in the second term could potentially be $A/2$ and where
$\epsilon$ represents the unknown (but formally universal) factors in
eq.~\eqref{eq:dcvAnswer}
\begin{equation}
  \label{eq:epsilon}
  \epsilon = \sum_{h} \cN_{h,-1} \frac{m_h^2}{Q_{0,h}} 
      \,\ln^{-A} \frac{Q_{0,h}}{\Lambda} \,.
\end{equation}
These results give us two distinct predictions.  Firstly
mass-dependent power corrections should have a leading piece which
goes as $(\ln^A Q/\Lambda)/Q$, where $A\simeq 1.565$ for $\nf = 5$.
Secondly the normalisation of this leading $Q$ dependence should be
predictable for all variables to within a new universal constant
factor $\epsilon$ which is intrinsically non-perturbative. There can
be additional corrections which are beyond our control, but their
scaling should be closer to that of a pure $1/Q$ term,
and therefore at very high energies they will be subleading.

\subsection{Absolute predictions}
\label{sec:qcdabs}

Within the tube model (and all renormalon based analyses) we had a
prediction that the leading hadronisation correction should scale as
$1/Q$. Yet from the arguments so far in this section we can see that,
even in the $E$-scheme, the differences between two different hadronic
levels (related by the decay of some hadron species) will involve a
term of order $(\ln^A Q)/Q$. Therefore for an arbitrary hadron level
the total hadronisation correction will also have a piece of order
$(\ln^A Q)/Q$.

One may well ask whether there exists a hadron level free of $(\ln^A
Q)/Q$ corrections. For example if one reconstructs the various
hadronic decays so as to arrive at the level of the `first hadronic
resonances created' then one is free of the corrections associated
with hadron decay. But it is difficult to define what is meant by the
first hadronic resonances, since one doesn't know how far `back' in
the decay chain to go, especially when one reaches resonances whose
width is of the same order as their mass: at this stage resonance
decay and the hadronisation process  become intertwined.

There are even reasons to believe that hadronisation itself could lead
to contributions of order $(\ln^A Q)/Q$, as is illustrated by the
following simplistic argument: in the same way that the decay of a
massive hadron (mass $m$, energy $E$) liberates a certain amount of
energy, roughly of order $m^2/2E$, in order to produce a massive
hadron one needs to \emph{supply} that amount of energy. The
reshuffling of momenta associated with `supplying this energy,' may
well affect the mean transverse momentum per unit rapidity.  After
summing over all hadrons, this implies a contribution to the mean
transverse momentum proportional to the $(-1)^\mathrm{th}$ moment of
the energy, \ie to $(\ln^A Q)/Q$.  Since the process at play should be
rapidity independent, we expect that for a particular event-shape
variable, $\cV$, the correction will be proportional to $c_\cV$ as was
the case for corrections due to hadron decay (and with the same
proviso concerning the broadenings). We point out that the change in
transverse momentum associated with the hadronisation could well be
negative, if the energy that is `reshuffled to produce the masses'
comes from transverse degrees of freedom.

So for a given hadronic level $\ell$, what we can say about the $(\ln^A
Q)/Q$ part of the hadronisation corrections, is that for a variable
$\cV$ in a scheme $S$ it has the form\footnote{In the case of the
  broadenings the full form actually has $4\gamma_{B_\decay}
  \mu_\ell$ rather than $c_\cV \mu_\ell$.}
\begin{equation}
  \label{eq:abslnQterm}
 \left( c_\cV \,\mu_\ell  \,+\, \gamma_{\cV_S} \,\epsilon_\ell \right) 
 \frac{\ln^A Q/\Lambda}{Q}\,,
\end{equation}
where $\epsilon_\ell$ is the same as $\epsilon$ defined in
section~\ref{sec:QCDappl}, but 
now specific to our hadronic level $\ell$; the scale $\mu_\ell$
relates to the (mass-dependent) change in mean transverse momentum per
unit rapidity coming from the hadronisation and subsequent decays of
resonances to our hadronic level $\ell$:
\begin{equation}
\label{eq:defmuell}
  \mu_\ell = \lim_{Q\to\infty} \frac{1}{\ln^A Q/\Lambda}
   \int \frac{d k_t}{k_t} k_t \left(
    \sum_h \Phi_h(k_t,0) - \sum_{p=q,g} \Phi_p(k_t,0) \right).
\end{equation}

We can make one further statement: there is phenomenological evidence
that our decay scheme gives a reasonable approximation to actual
hadronic decays, or more specifically that regardless of the initial
hadronic level, the decay-scheme results are almost identical (\cf
section~\ref{sec:reshaddec}). Accordingly, for any pair of hadronic
levels $\ell$ and $\ell'$ we expect the following relation to hold
\begin{equation}
  \label{eq:muplusepsilon}
  4 \mu_\ell  + \epsilon_\ell \simeq 4 \mu_{\ell'}  + \epsilon_{\ell'}\,,
\end{equation}
where for variables other than the broadening we have exploited the
fact that $4\gamma_{\cV_{\decay}} = c_\cV$.

\subsection{Infrared and collinear saftey?}

All the event-shapes considered in this paper are generally considered
to be infrared and collinear (IRC) safe. Yet above we have argued that
they are sensitive to hadron multiplicities which are inherently IRC
\emph{unsafe}. How can these two statements be reconciled?

Event-shapes are perturbatively IRC safe because they are linear in
particle momenta; so if a parton of energy $E_0$ splits collinearly
into two partons of energies $E_1$ and $E_2$ then the value of the
event shape is unchanged, $\cV \sim E_1/Q + E_2/Q = E_0/Q$. 

Mass effects behave differently because, simply kinematically, they
are proportional to $m^2/EQ$. They are not usually to be seen in
perturbative calculations because 
partons are considered to be massless.
But hadrons are
massive and since energy is now in the denominator, the relation
$1/E_0 \neq 1/E_1 + 1/E_2$ means that mass effects appear to be IRC
sensitive.  
There are also situations where one would expect multiplicity
enhancements, similar to those discussed here, in purely perturbative
calculations. With massive $b$ quarks for example, in the difference
between $p$ and $E$ schemes one would see an $m_b/Q$ suppressed
contribution. This would be sensitive to the multiplicity of slow
large-angle $b$ quarks, which at high orders is be enhanced by
infrared and collinear logarithms of $m_b/Q$.

But mass effects do not turn an IRC safe observable into an IRC unsafe
one. They are always suppressed by powers of $m/Q$, so even if this
factor is enhanced by logarithms of $m/Q$, in the limit $Q\to \infty$
the net contribution still goes to zero, as required for IRC safety.

\section{Comparison to Monte Carlo}
\label{sec:MC}

It would be nice to test our predictions of mass effects against data,
for example by looking at the differences between measurements of the
same variable in two different schemes.  The experimental difficulties
are significant though.

To calculate event shape observables in an experiment, four-momenta
have to be reconstructed from the tracking and calorimetric data.  As
a simultaneous measurement of $p$ and $E$ is far too imprecise to
constrain the mass, one usually explicitly assigns a mass to each of the
reconstructed particles.  The mass assignment is based on the
signature in the detector. In general it allows the separation of
neutral from charged and electromagnetic from hadronic particles, but
the separation of different hadrons ($\pi$, K, p or n) is
experimentally much more difficult and usually left to specialised
tagging algorithms.  Often the pion mass is chosen to be assigned to
charged hadrons, as pions are the most common charged hadrons. For
neutral particles, in principle calorimetry could distinguish between
electromagnetic and hadronic particles, but in practice this is
difficult and all neutrals are assigned zero mass.

In any case the
effect of misassignment needs to be corrected using Monte Carlo
simulation.  This means that `measurements' of differences between
schemes depend critically on the extent to which the Monte Carlo
simulator gives an accurate description of features of the data which
are not measured. Of course the Monte Carlo programs have themselves
been tuned in order to reproduce the (separate) data on the production
of different hadron species, but this tuning does not constrain all
the available degrees of freedom. As a result it is difficult to
establish the magnitude of those systematic errors on the measurement
that are associated with the dependence on the particular Monte Carlo
model that has been used for calculating corrections.

\subsection{Magnitude of mass effects}
\label{sec:MCissues}

When comparing the results from Monte Carlo simulations with data one
cross-check comes from comparing the absolute value of an observable.
Table~\ref{tab:delta_rho_data} shows the invariant jet mass calculated
in different schemes from Monte Carlo as well as from data. Within the
experimental errors the simulations agree with the experimental
results. The difference between the schemes, though, is more sensitive
to the choice of the Monte Carlo program. Different simulations
deviate up to $\pm10\%$ indicating a systematic uncertainty of that
order. The differences computed from the DELPHI results confirm that
the simulation used by DELPHI gives consistent results.

\begin{table}[bp]
  \begin{center}
    \begin{tabular}{|c|c|c|c|c|c|}
     \hline
                  & Herwig 6.1 & \multicolumn{2}{c|}{Pythia 6.1}& Ariadne  & DELPHI\\ 
\raisebox{1.2ex}[0pt][0pt]{Observable}& default &default   & tuned             & default & data \\ 
     \hline
     $\rho$       & $0.0363$ & $0.0372$  & $0.0371$          & $0.0375$  & $0.0370 \pm 0.0005$ \\
     $\rho_{p}$ & $0.0316$ & $0.0326$  & $0.0332$          & $0.0330$  & $0.0327 \pm 0.0003$ \\
     $\rho_{E}$ & $0.0326$ & $0.0336$  & $0.0341$          & $0.0340$  & $0.0335 \pm 0.0003$ \\
     \hline
     $\delta\rho_{p}$ & $0.0047$ & $0.0046$  & $0.0039$          & $0.0045$  & $0.0043$     \\
     $\delta\rho_{E}$ & $0.0037$ & $0.0036$  & $0.0030$          & $0.0035$  & $0.0035$ \\
     \hline
    \end{tabular}
  \end{center}
\caption{ 
     Comparison of MC with data for standard and $p$-scheme at $91.2\GeV$.
     The statistical errors on the MC results are below 0.0001. Differences
     between the different models indicate systematic uncertainties, which for
     $\delta\rho_{p}$ are of the order of 10\%. 
     DELPHI numbers were obtained by averaging the published \cite{Delphi}
     results for heavy and light-jet masses. The tuning used for the
     tuned Pythia results is based on  \cite{tuning}.}
   \label{tab:delta_rho_data}
\end{table}

\begin{figure}[tbp]
  \begin{center}
    \epsfig{file=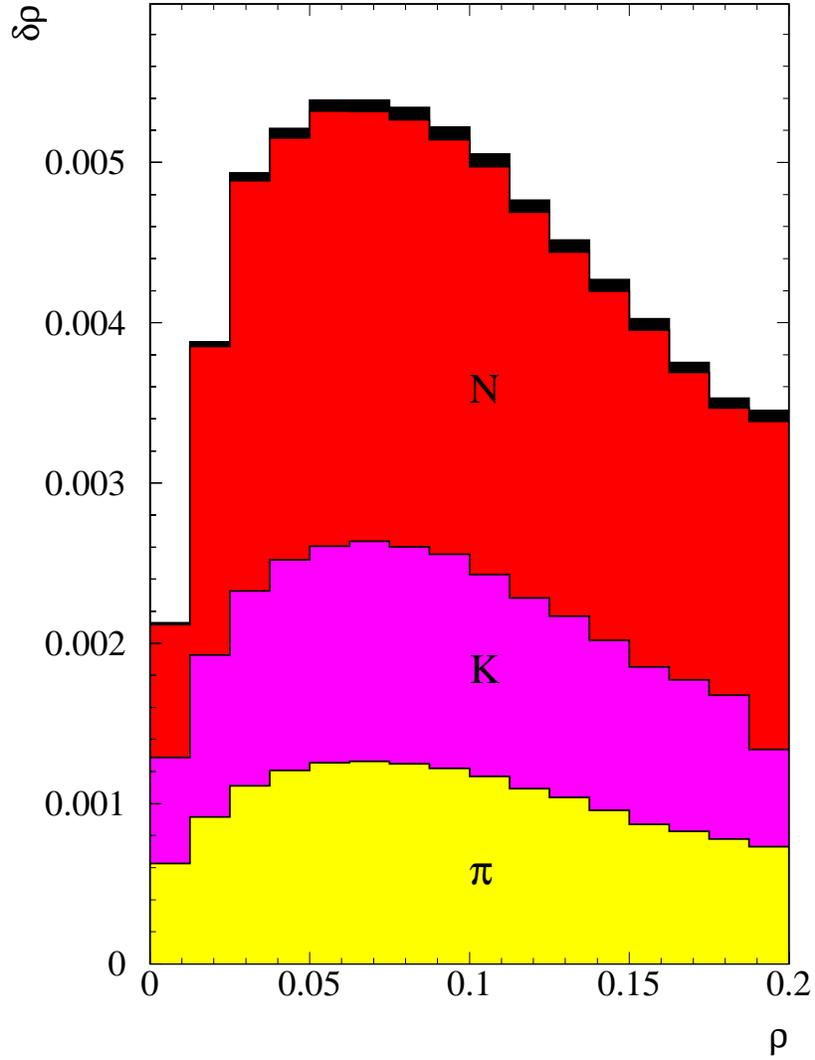,width=0.7\textwidth}
    \caption{The contributions to $\delta \rho$, the difference between 
      the mean jet mass $\rho$ in the default (massive) and
      $p$-scheme, coming from different hadron species ($\pi^{\pm}$,
      kaons and nucleons), shown as a function of $\rho$ in the
      default scheme.  }
    \label{fig:PiRhoN}
  \end{center}
\end{figure}

It is also interesting to see how the shift caused by switching from
default to $p$-scheme depends on the value of the observable and how the
different particle species contribute.  Figure~\ref{fig:PiRhoN}, 
for the jet mass, was obtained using Ariadne~\cite{Ariadne}
separating the contribution from different particle species by
applying the $p$-scheme only to particles of a given type and stacking
the results.

Because the particle masses enter quadratically into $\rho$
nucleons give the biggest effect, despite their low multiplicity.
At low values of $\rho$ changing the scheme causes a low average shift,
which first rises with $\rho$ and then falls off again.
This structure stems from all particle types. 
An additional peak at large values of $\rho$ is due to nucleons only and
probably arises due to peculiarities of baryon production in 4-jet events.

The overall structure
reflects influences from the numerator and the denominator in
the definition of $\rho$. Low values of the numerator are probable only at low
multiplicities which in turn only have small corrections. With larger values
of $\rho$ the average multiplicity rises and so does the average
difference between standard
and $p$-scheme. For even larger values the relative change in the numerator
decreases and mass-effects in the denominator get progressively more
important, finally  over-compensating the changes in the numerator. 

The interplay between numerator and denominator results in a complex
$\rho$-dependence of the difference between schemes and demonstrates
that mass-effects have a non-trivial influence on the observables shape.
Mean values are, however, dominated by small values of the numerator, where
the influence of the denominator is suppressed 
(see section~\ref{para:suppressed_denominator}).

While the necessary experimental corrections make it difficult to
measure the difference between the schemes from data, the experiments'
reliance on simulation allows us to correct the existing data to any
desired scheme without introducing significant additional systematic
errors.  We shall use Ariadne~\cite{Ariadne} to transform existing
data~\cite{\alldata} to a desired scheme whenever corresponding
measurements in this scheme do not exist. Currently data exists for
non-standard schemes only from DELPHI \cite{Delphi}, H1 and ZEUS
\cite{HERApscheme}.

\begin{figure}[tbp]
  \begin{center}
    \epsfig{file=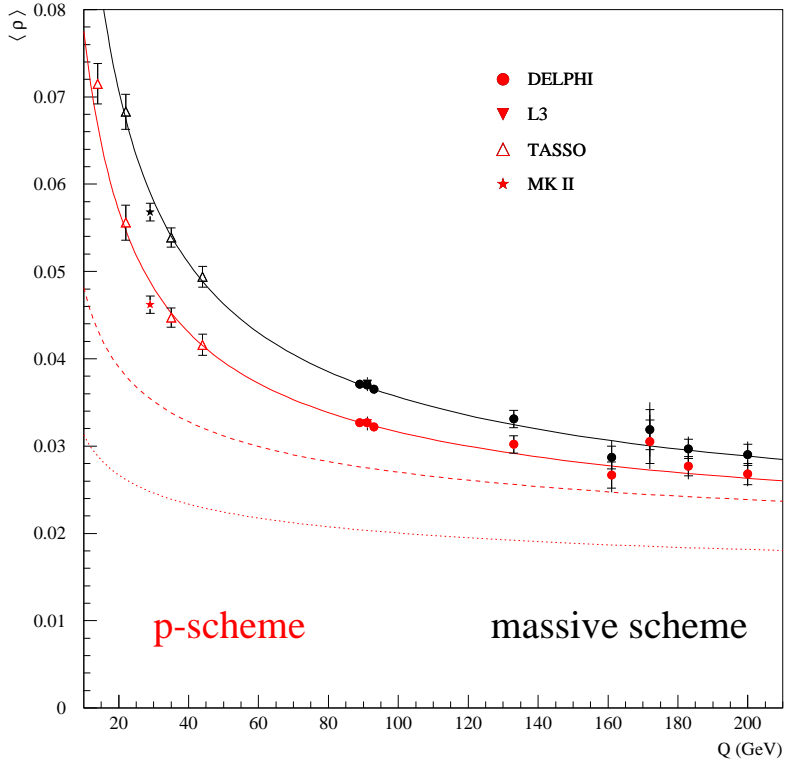,width=0.7\textwidth}
    \caption{The jet mass $\rho$ as measured in the default scheme
      (upper set of points) and corrected to the $p$-scheme (lower set
      of points). The upper curve corresponds to a fit to the
      default-scheme points, while the lower three curves correspond
      to a fit to the $p$-scheme points (the dotted curve is the pure
      $\as$ component, the dashed curve includes the $\order{\as^2}$
      term while the solid curve includes a power correction of the
      form eq.~\eqref{eq:a0Toalpha0}).}
    \label{fig:rho}
  \end{center}
\end{figure}

In figure~\ref{fig:rho} we show data for the jet mass $\rho$ as
function of $Q$ in the default and in the $p$-schemes.  As $\rho$ is
usually not given by the experiments, it was taken as the average of
the measured heavy and light-jet masses.  The lines in
figure~\ref{fig:rho} correspond to fits using the $\cO{\as^2}$
perturbative prediction and a power correction of the form
eq.~\eqref{eq:a0Toalpha0}.

The difference between the two middle (red solid and dashed) curves
corresponds to the normal, `universal', power correction. The
difference between the upper two (red and black solid) curves is the
mass-dependent power correction.  One sees that above $Q=M_Z$ it is as
large as the traditional $1/Q$ term, and it can be as much as $10\%$
of the mean value of the observable (\cf also
table~\ref{tab:delta_rho_data}).  For a general observables
differences between the $p$ and $E$-schemes at $Q=M_Z$ are of the
order of a few percent, whereas differences between decay and
$E$-schemes are between $5$ and $10\%$ of the observable.

So while none of our previous analysis has given us any indication of
the absolute size of mass-induced effects, comparisons in this section
have shown that the absolute size of mass effects at $Q=M_Z$ due to
hadrons can be sizable portion of the non-perturbative
power-corrections.

Further results of comparisons with data, transformed to various
schemes, will be presented in section~\ref{sec:data}.

\subsection{Comparison to predictions}
\label{sec:MCcomparison}

\begin{figure}[tbp]
  \begin{center}
    \epsfig{file=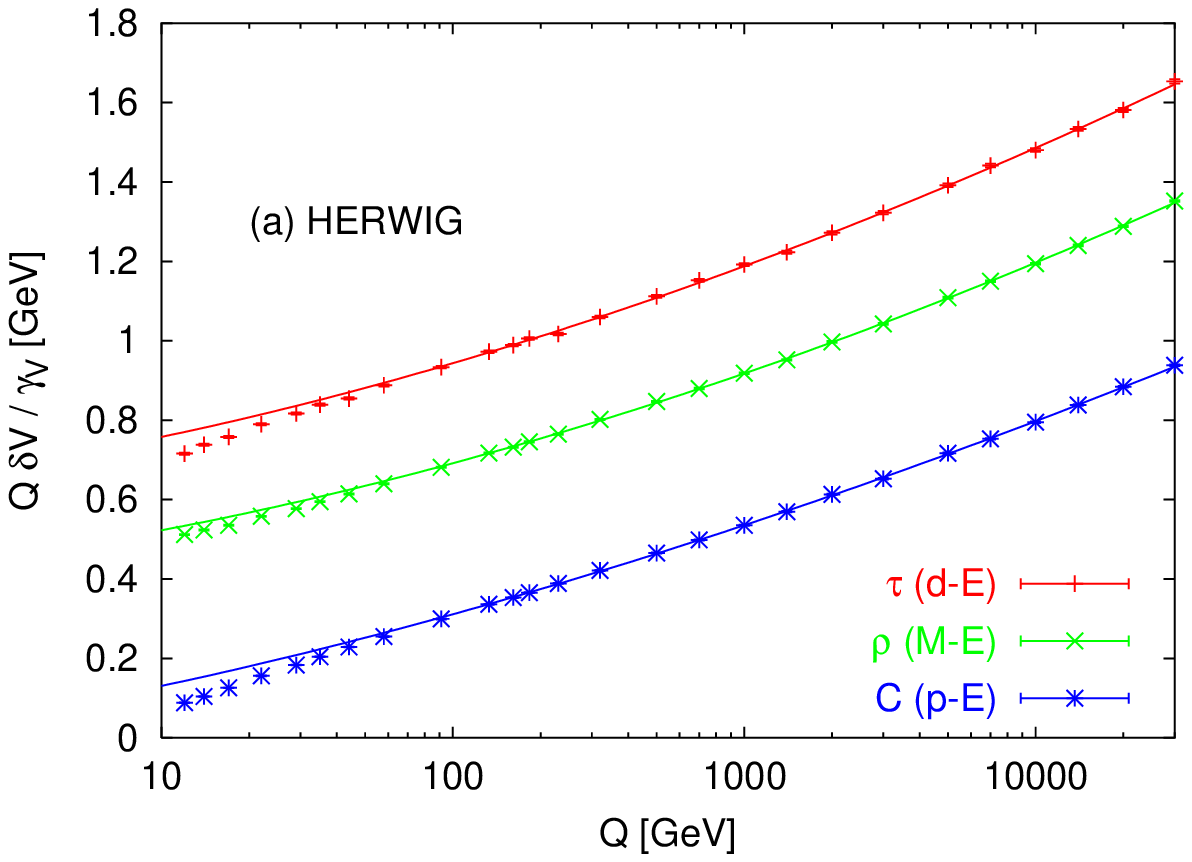,width=0.59\textwidth}
    \epsfig{file=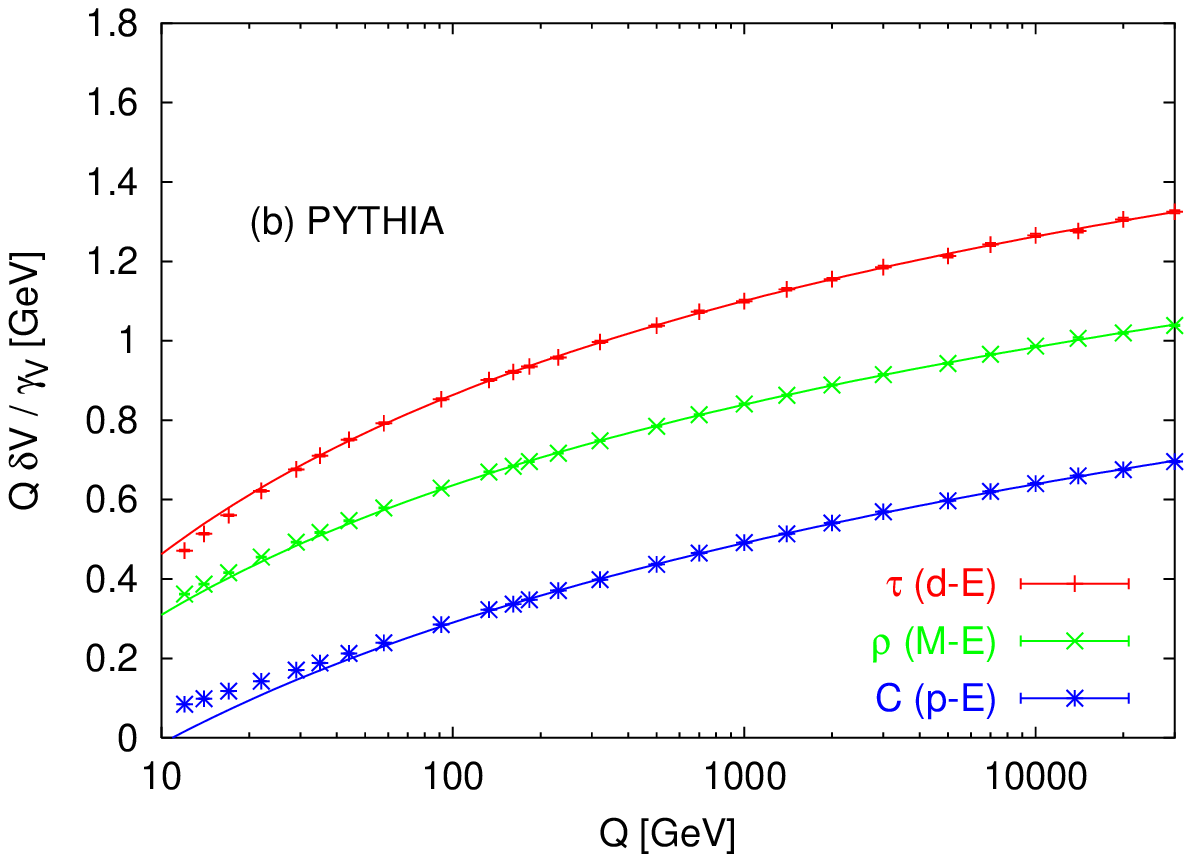,width=0.59\textwidth}
    \epsfig{file=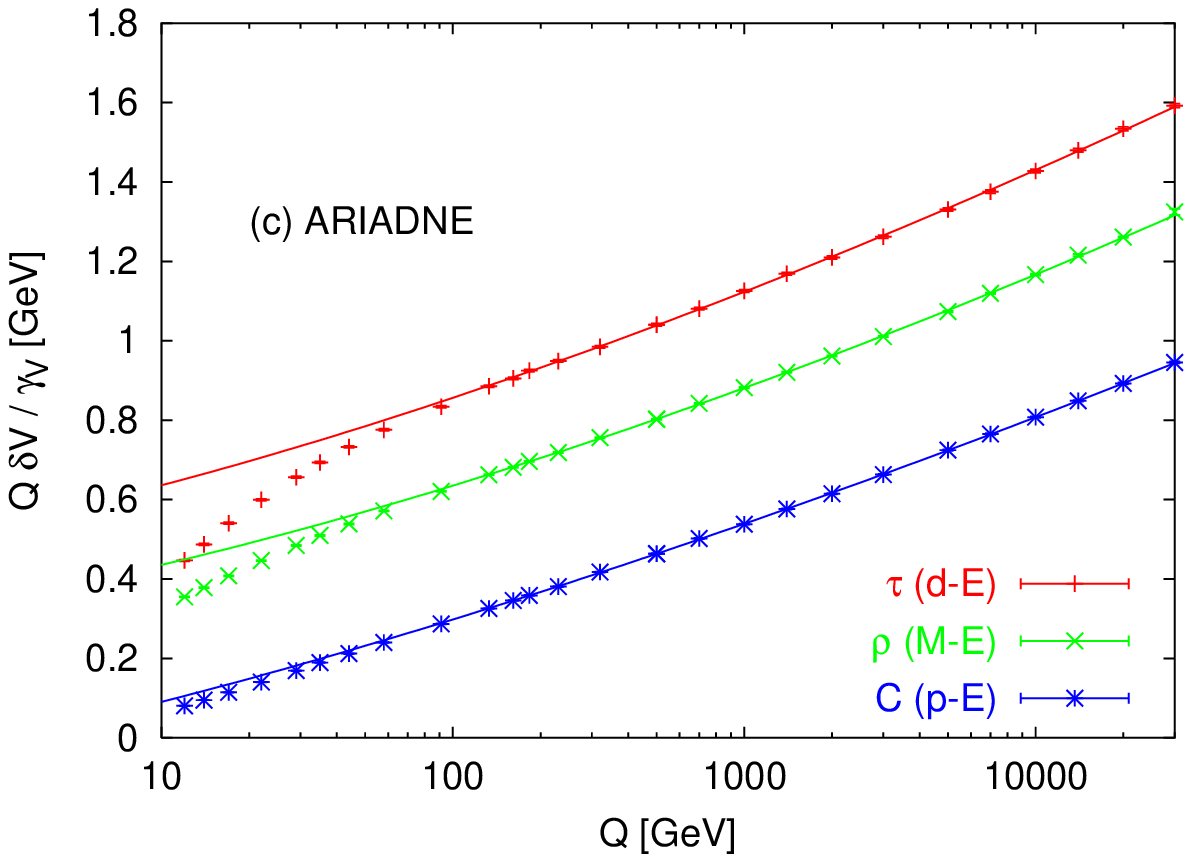,width=0.59\textwidth}
    \caption{Herwig, 
      Pythia 
      and Ariadne 
      results for the differences between pairs of measurement schemes
      for three observables. In all cases the differences have been
      normalised to $\gamma_\cV$. The lines are fits of the form
      eq.~\eqref{eq:fitformula} to the points with $Q>100\GeV$. These
      figures have been generated using events with primary
      down-quarks; the labels $(d-E)$, $(M-E)$ and $(p-E)$ indicate
      differences between the decay and E-schemes, the massive and
      $E$-schemes and the $p$ and $E$-schemes respectively.}
    \label{fig:HerwigvPythia}
  \end{center}
\end{figure}

Here we compare Monte Carlo results with the predictions of
sections~\ref{sec:tube} and \ref{sec:QCD}. There are two main
predictions which we wish to test. Firstly that the leading
energy-dependence of mass-dependent effects is $(\ln^AQ)/Q$; secondly
that the coefficient of the leading energy dependence is proportional
to $\gamma_\cV$. We are also interested in examining a third, more
qualitative prediction, namely that certain subsets of observables
have similar subleading $1/Q$ mass effects.
 
We shall study results from three Monte Carlo event generators: Herwig
\cite{Herwig}, Pythia \cite{Pythia} and Ariadne \cite{Ariadne}. Let us
first illustrate the kind of behaviour that is seen by examining three
observables: the difference between the $p$ (default) and $E$-schemes
for the $C$-parameter, the difference between the massive (default)
and $E$-schemes for the jet mass and the difference between the decay
and $E$-schemes for the thrust. These differences (multiplied by $Q$)
are shown as a function of $Q$ in figures~\ref{fig:HerwigvPythia}a,
\ref{fig:HerwigvPythia}b and \ref{fig:HerwigvPythia}c for Herwig and
Pythia and Ariadne respectively.\footnote{The plots have been
  generated using only events with primary down-quarks: there seem to
  be slight differences between the results coming from different
  light-quark species and the change in flavour composition of events
  as $Q$ approaches $M_Z$ leads to an extra small but spurious
  $Q$-dependence. More important though is the removal of top-quark
  production: for $Q \gtrsim 2m_t$ the structure of the Born level of
  primary top-quark events is very different from that primary
  light-quark events, because of the top decay.} They have all been
normalised to the appropriate $\gamma_\cV$.

Pure $\gamma_\cV/Q$ corrections would lead to superimposed flat lines.
The fact that the lines rise for all three event generators is
consistent with the fact that we have a correction enhanced at larger
values of $Q$. But the nature of the $Q$ dependence is not consistent
between the different programs. For Herwig and Ariadne the second
derivative is positive and roughly consistent with $\ln^A Q$ with $A
\simeq 1.6$ as predicted in eq.~\eqref{eq:Final}. On the other hand
the Pythia results are inconsistent with such a hypothesis --- we
return to this problem shortly.


Our second prediction in eq.~\eqref{eq:Final} was that the
$Q$-dependence should be proportional to $\gamma_\cV$ --- the fact
that our observables (normalised to $\gamma_\cV$) have very similar
$Q$ dependences supports this hypothesis. This is true for all three
event generators.


To study these questions more systematically we fit a formula of the
following form\footnote{In
  the case of the broadenings in the decay scheme we actually use a
  more complicated form, in line with the discussion in
  appendix~\ref{sec:broad}:
\begin{equation*}
  \langle \delta_m \cV \rangle =
  \frac1Q
  \left(\gamma_{\cV,\om=1}  \epsilon \ln^{A_{\eff} }
    \frac{Q}{\Lambda} \,+\, \gamma_{\cV,\om=\infty} B  \right).
\end{equation*}
}
\begin{equation}
  \label{eq:fitformula}
  \langle \delta_m \cV \rangle = 
  \frac{\gamma_\cV}Q \left(\epsilon \ln^{A_{\eff}}
    \frac{Q}{\Lambda} \,+\, B \right) \,,
\end{equation}
to $\langle \delta_m \cV(Q)\rangle$, where $\epsilon$, $A_{\eff}$ and
$B$ are the fit parameters (we take $\Lambda=0.2\GeV$).  Implicit in
this procedure is the assumption that terms with subleading energy
dependence are reasonably well approximated by $\gamma_\cV B/Q$ (fits
involving more sophisticated forms for the subleading terms turn out
to be fairly unstable). To reduce the impact of subleading effects we
only fit points with $Q > 100\GeV$. We have generated $10^5$ events
per point.

\begin{figure}[tbp]
  \begin{center}
    \epsfig{file=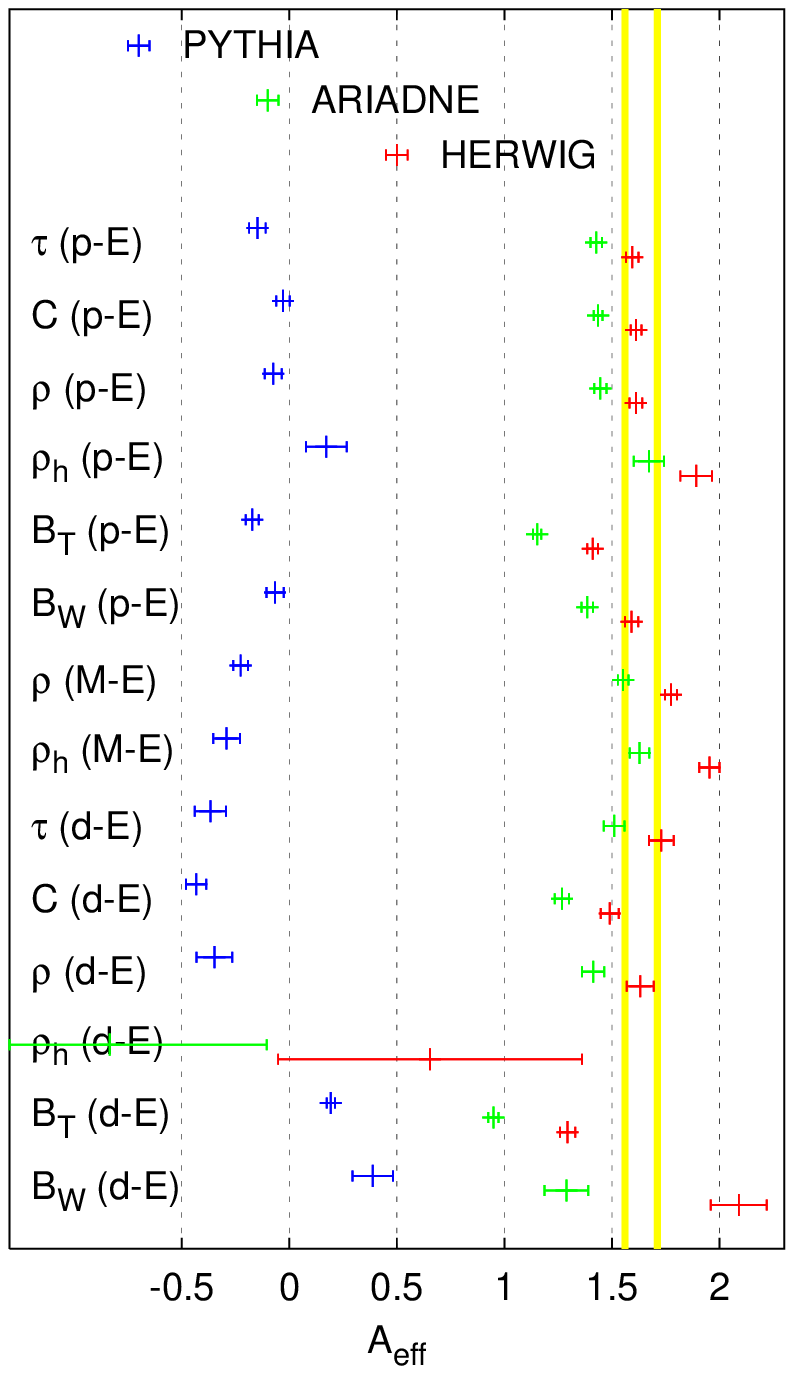,width=0.48\textwidth}\;\;
    \epsfig{file=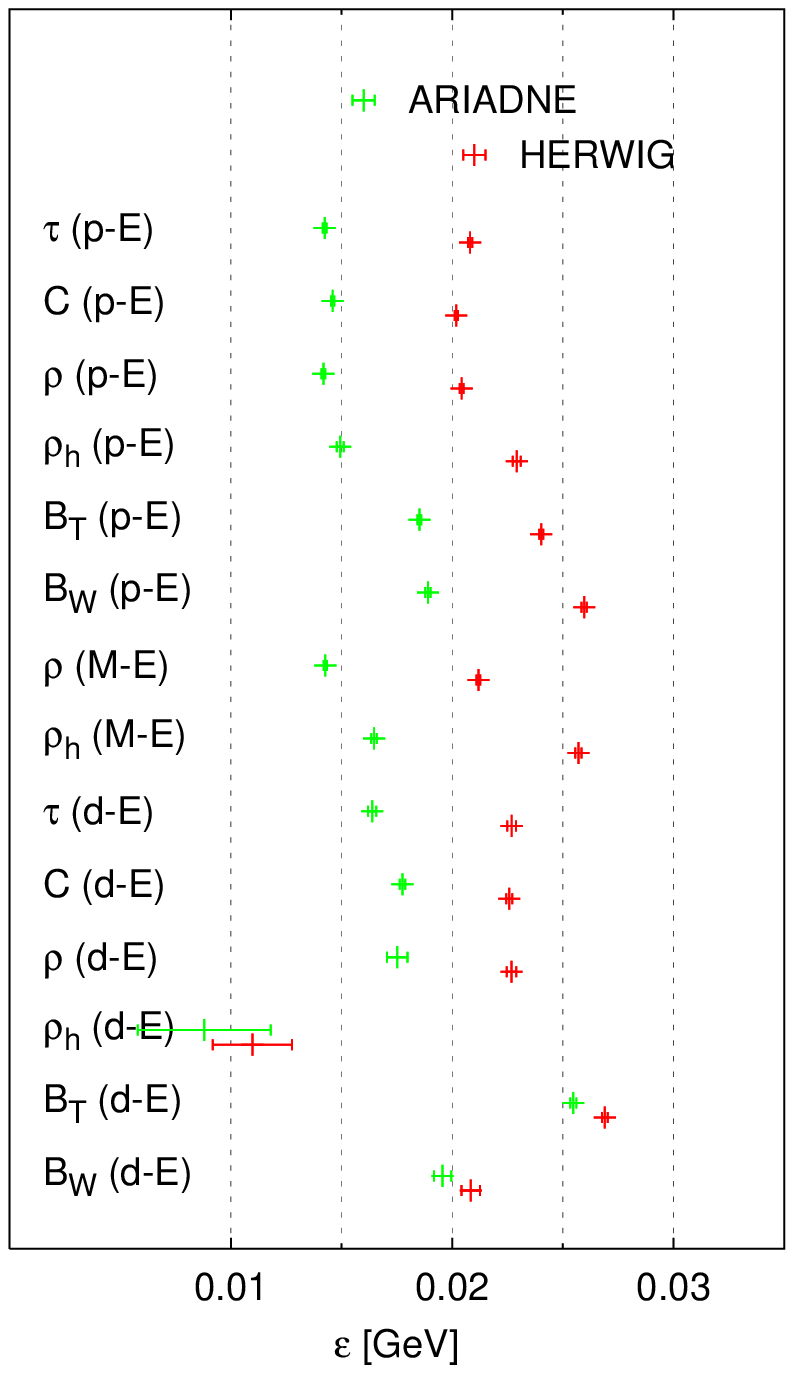,width=0.48\textwidth}
    \caption{The left-hand plot shows the power $A_{\eff}$ `measured' for
      a range of observables from Pythia, Ariadne and Herwig; the
      yellow bands are the expected values for $\nf=5$ and $\nf=6$.
      With $A_{\eff}$ fixed to its predicted $\nf=5$ value, the values
      of $\epsilon$ fitted for the different observables are shown in
      the right-hand plot (just Ariadne and Herwig). The fits are
      carried out for $Q> 100\GeV$ to reduce their sensitivity to
      subleading effects using only events with $d$ primary quarks.
      Errors are statistical from $10^5$ events at each of
      17 energies.}
    \label{fig:mainresults}
  \end{center}
\end{figure}

We expect $A_{\eff}$ to be somewhere between the $\nf=5$ and $\nf=6$
values of $A=1.565$ and $A=1.714$. The results for $A_\eff$ are shown
in the left hand plot of figure~\ref{fig:mainresults}, for all
three event generators, together with bands representing the predicted
$\nf=5$ and $\nf=6$ values for $A$. Almost all the Pythia results have
$A_\eff\simeq0$ which is a signal of a dependence of the form $\ln \ln
Q/\Lambda$. The Herwig and Ariadne results in general lie close to the
predicted value for $A$. Most of the observables lie to within
$\pm0.1$ of the expected range, the notable exceptions being the
difference between the decay and 
$E$-schemes for the heavy-jet mass and the broadenings. The
broadenings in the decay scheme are quite delicate 
observables because they involve an expansion in powers of $\sqrt{\as}$
which may be quite slowly convergent. In the case of the heavy-jet
mass it is not too clear what is going wrong though it may well be
related to the non-inclusiveness of the variable (\cf
appendix~\ref{sec:mh}).

It is of interest to establish whether the inconsistency between our
prediction and Pythia is due to the nature of the hadronisation or to the
parton showering. If it were the former we might think that we had
been too naive in assuming LPHD and an absence of qualitative changes
due to hadronisation. Two distinct arguments support the hypothesis
that the problem lies with the parton showering. One is that Ariadne
uses the same string fragmentation routines as Pythia and so the
problem must lie in the part of the physics which is treated
differently between the two programs (the parton showering). The
second argument comes from a direct investigation of the Pythia parton
level: in Pythia, parton level gluons are massless so we cannot simply
look at the difference between two mass-schemes. However mass effects
are just related to the sum of the inverse energies of all the
particles, so we can instead look directly at the behaviour of
$D_{-1}(Y = \ln Q/\Lambda)$ (\cf eq.~\eqref{eq:MomMult}) at both
parton and hadron level (where it is best examined for individual
hadron species) and check that it has the right energy dependence. We
find that in Pythia both at parton and hadron levels, $D_{-1}(Y)$
rises too slowly with energy, while Herwig for example shows an energy
dependence which is consistent with our prediction, both at parton and
hadron level. This suggests that the parton showering present in
Pythia might be lacking some of the dynamics associated with the
coherent branching approach used in section~\ref{sec:QCD}. 

Problems with coherence should have implications also for hadron
multiplicities. If we restrict ourselves to `uds' primary-quark
events, we find that the ratio of Herwig and Ariadne $\pi^{\pm}$
multiplicities is essentially independent of $Q$ (to within $1\%$ for
$Q$ between $100\GeV$ and $30\TeV$). The ratio of Pythia to Ariadne
$\pi^{\pm}$ multiplicities on the other hand decreases by about $13\%$
over this range. We note in passing that the $\pi^{\pm}$
multiplicities from Herwig and Ariadne at any given $Q$ differ by
about $8\%$, and that if one includes heavy primary quarks the
situation is more complicated.

\begin{figure}[htbp]
  \begin{center}
    \epsfig{file=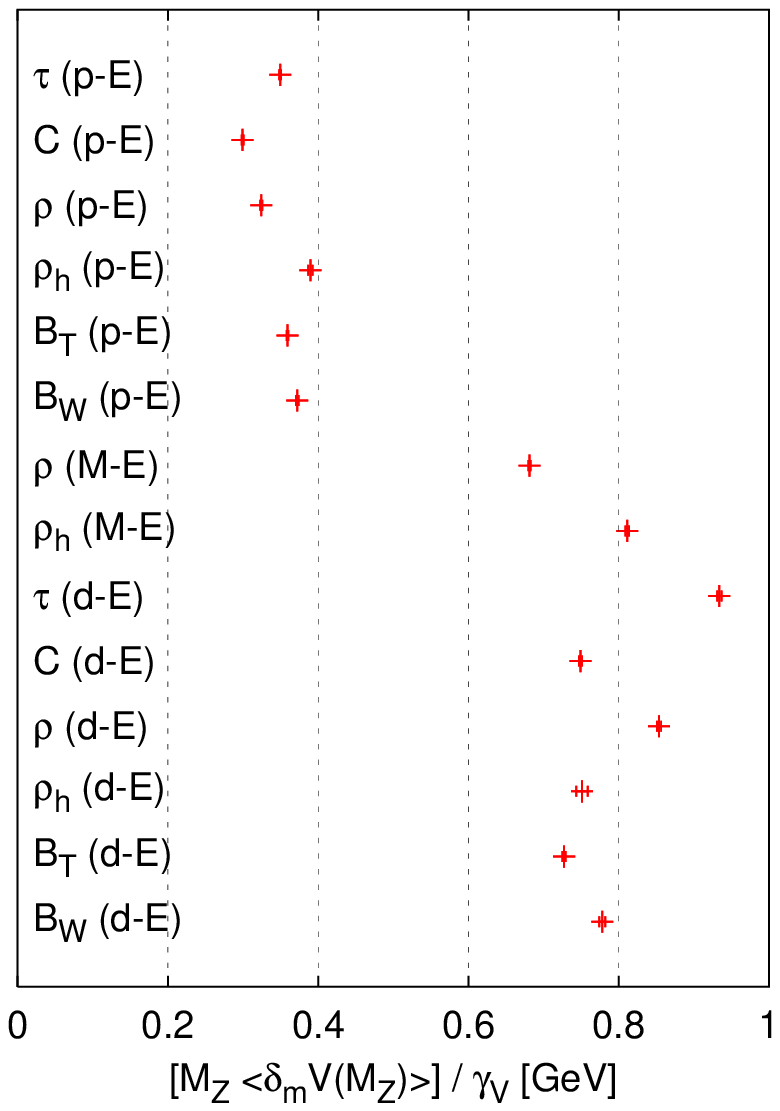,width=0.48438\textwidth}\;\;
    \epsfig{file=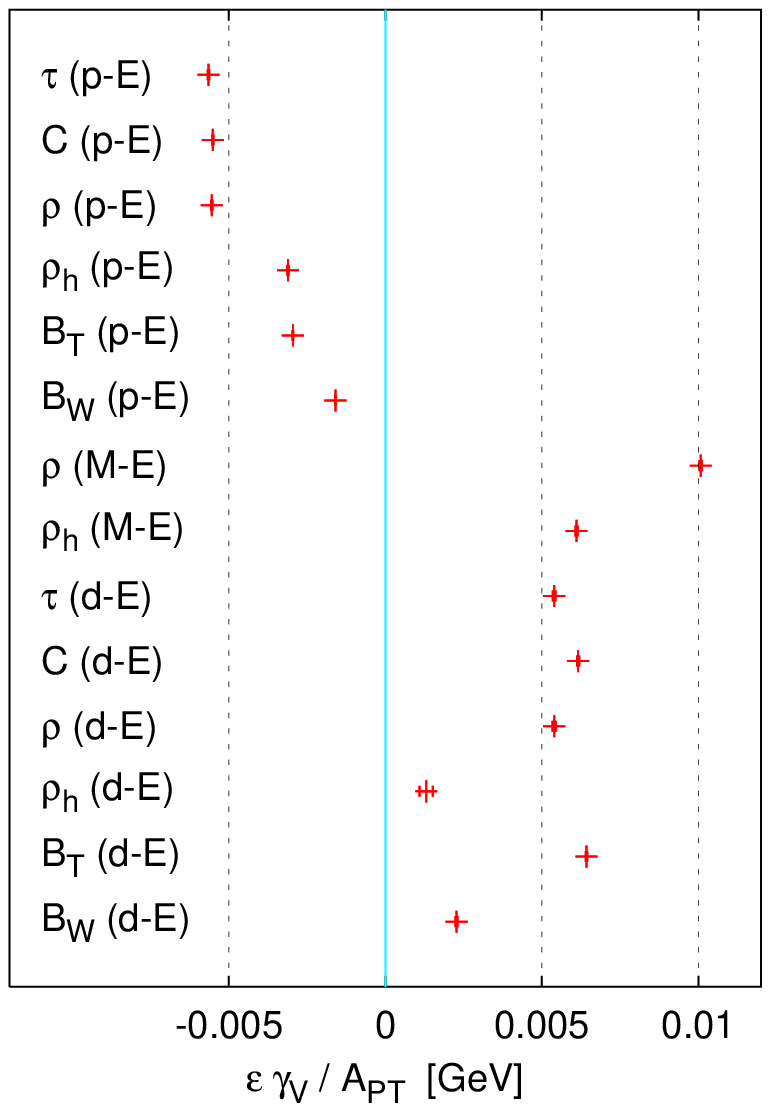,width=0.48\textwidth}
    \caption{The left hand plot
      shows the value of the observables at a fixed value of $Q=M_Z$,
      normalised to $\gamma_\cV$. The right hand plot shows the
      $Q$-dependence of the various observables normalised to the
      leading-order perturbative coefficient rather than to
      $\gamma_\cV$. Results are shown for Herwig only.}
    \label{fig:subsresults}
  \end{center}
\end{figure}

Our second prediction was that the leading $Q$-dependence should be
proportional to $\gamma_\cV$. To test this systematically we fix
$A_\eff$ to be equal to $A=1.565$ and fit for the values of $\epsilon$
and $B$ in eq.~\eqref{eq:fitformula} (using the same range of $Q$ as
before). We do this only for Herwig and Ariadne. The results are shown
in the right hand plot of fig.~\ref{fig:mainresults}. The mean value
of $\epsilon$ is about $0.023\GeV$ for Herwig and about $0.016\GeV$
for Ariadne. One should not be misled into thinking that these small
numbers imply small effects --- they get multiplied by $\ln^A
Q/\Lambda$, which is about $17$ for $Q=M_Z$! The range of $\epsilon$
values for different observables is typically about $\pm 10\%$ from
the central value (with the exception of the difference between the
decay and $E$-schemes for $\rho_h$). Thus our two main predictions,
concerning the energy dependence and the relative normalisation of
mass-dependent effects are in remarkable agreement with Monte Carlo
results.

We can also examine the mass correction at a given fixed
value of $Q$, rather than its $Q$-dependence. In the left-hand plot of
figure~\ref{fig:subsresults} we show $Q\langle \delta_m \cV(Q)\rangle
/ \gamma_\cV $ for $Q=M_Z$. The points seem to fall into two groups:
those corresponding to differences between the $p$ and $E$-schemes,
and those corresponding to differences between the massive and
$E$-schemes (jet masses) and the decay and $E$ schemes.  The
differences between and $p$ and $E$-schemes are all governed by the
functions $\delta c_\cV$ in figure~\ref{fig:DeltaCV} which have a
maximum: these functions have very similar shapes, meaning that
whatever the form of $\Phi_h(k_t,Y)$ in \eqref{eq:fullDeltaMV} the
integral will be proportional to $\gamma_\cV$. This statement was made
earlier in the form of eq.~\eqref{eq:ratioNonUniv}. The fact that the
points for the jet masses (massive minus $E$-schemes) and the decay
minus $E$-schemes are higher was anticipated in
eq.~\eqref{eq:ratioNonUnivRho}, since after accounting for the
rescaling by $\gamma_\cV$, their $\delta c_\cV$'s remain larger than
those of the other variables.

Finally, for entertainment purposes, in the right hand plot of
figure~\ref{fig:subsresults} we show the $Q$-dependence normalised not
to $\gamma_\cV$ but to $A_{PT}$, the first coefficient of the
perturbative expansion of the event shape. It illustrates the fact
that mass effects vary significantly in size and sign from one
observable to the next --- and that the ability to predict that
pattern of this variation is a non-trivial achievement!

\subsection{Resonance and hadron-level decay scheme results}
\label{sec:reshaddec}

In section~\ref{sec:decayscheme} we mentioned that a phenomenological
advantage of the decay scheme is that regardless of the hadronic level
from which we start the decay scheme results are very similar. To
verify this we start with two hadronic levels: a normal hadronic
level, as defined earlier and a resonance level which is `defined' as
the first level of hadrons produced in Pythia or Ariadne. We then look
at the difference between decay-scheme event-shape values for these
two hadronic levels compared to the difference between the $E$-scheme
values. Results obtained from Ariadne are given (in $\%$) in
table~\ref{tab:decay}: they show that for most variables, in the decay
scheme one is very insensitive to the initial hadronic level. This is
not completely trivial since the actual decays that take one from
resonance to hadron level are not just the artificial massless
two-body decays of the decay scheme.

\begin{table}[b]
  \begin{center}
    \begin{tabular}{|c|c|c|c|c|c|c|}
      \hline
      $\cV$ & $\tau$ & $\rho$ & $\rho_h$ & $C$  & $B_T$ & $B_W$    \\
      \hline
      $\frac{\cV_{\decay,\had} - \cV_{\decay,\res}}{\cV_{E,\had} -
        \cV_{E,\res}} (\%)$ 
      &   \footnotesize$1.1\pm0.2$ & \footnotesize $-7.3\pm0.9$&
      \footnotesize $0.8\pm0.2$ & 
      \footnotesize $1.9\pm0.1$ & \footnotesize $1.5\pm0.2$ &
      \footnotesize $-1.5\pm0.5$ \\ \hline  
    \end{tabular}
    \caption{Percentage dependence of decay scheme results on the
      choice of the initial hadronic level (`normal hadron' or
      resonance level), relative to the dependence of $E$-scheme
      results on the hadronic level. The numbers are shown for $Q=M_Z$
      using results from Ariadne. }
    \label{tab:decay}
  \end{center}
\end{table}

For most of the variables the ratio shown in table~\ref{tab:decay}
scales roughly as $1/Q$, \ie at these energies the difference between
decay schemes for the two hadronic levels is dominated by a $1/Q^2$
correction rather than a $1/Q$ correction. The jet mass $\rho$ and the
wide-jet broadening seem to be more complex (note also the different
sign of the correction compared to the other observables, and the
somewhat larger value for $\rho$), but the origins of the differences
have yet to be identified.

A point worth noting (we will see a related point in
section~\ref{sec:data}) is that at lower energies, the good
correspondence between the two decay-scheme results holds only if
heavy-quark decays are taken into account separately.

\subsection{Total hadronisation}

So far in this section we have examined differences between various
measurement schemes and various hadronic levels. We observed in
section~\ref{sec:qcdabs} that, since the difference between any two
hadronic levels contains terms proportional to $(\ln Q)^A/Q$, the
total hadronisation corrections in going to an arbitrary hadronic
level must also contain such terms. Additionally, hadronisation itself
might introduce a contribution proportional $(\ln Q)^A/Q$, as a
consequence of the reshuffling of momenta associated with the
production of massive hadrons. We introduced the parameter $\mu_\ell$
to represent the normalisation of such a component for a given
hadronic level $\ell$, and pointed out that it could quite conceivably
be negative, for example if the energy required to produce the hadron
masses comes partially from transverse degrees of freedom.

\begin{figure}[tb]
  \begin{center}
    \epsfig{file=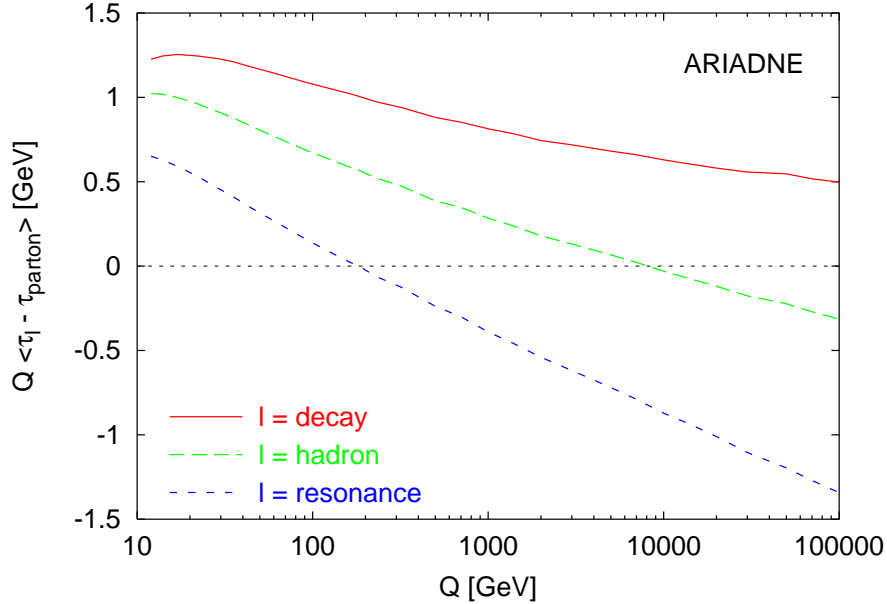,width=0.7\textwidth}
    \caption{The difference between hadron and parton level results
      for $\tau$ (multiplied by $Q$), shown as a function of $Q$.
      Curves are shown for three different hadronic levels, and have
      been obtained from the Ariadne event generator with only light
      primary quarks. The thrust is defined in the $E$-scheme.}
    \label{fig:totalhadr}
  \end{center}
\end{figure}

One may wonder what happens in the hadronisation models used in Monte
Carlo event generators.  The various curves in
figure~\ref{fig:totalhadr} show the corrections to $\tau$ (multiplied
by $Q$) in going from parton level to each of a variety of hadronic
levels, as determined from Ariadne. The first level after
hadronisation is the resonance level --- the fact the corresponding
line has negative slope means that $\mu_\ell$ is negative. Indeed the
hadronisation corrections change sign at around $200\GeV$!

However we expected mass effects to have a characteristic signature,
namely to contain a term $(\ln Q)^A/Q$ with $A\simeq 1.6$. If we carry
out a fit (analogous to those carried out in
section~\ref{sec:MCcomparison}) to determine the effective power for
the resonance level curve then we obtain $A_\eff \simeq 1.0$. This
could be due to subleading mass effects, or some completely different
effect which has yet to be considered.

At other hadronic levels one sees a weaker $Q$ dependence --- the
positive $(\ln Q)^A/Q$ contribution from hadron decays cancels a large
part of the negative contribution from the hadronisation. This in
itself is an interesting, and even quite natural, result: the negative
contribution from hadronisation is of the same order as the positive
contribution from the decay of all hadrons!

Of course these observations may be very specific to the hadronisation
model being considered. It would have been interesting to carry out a
similar exercise with the Herwig event generator, however there the
situation is more complex because the gluons have a (large) mass
($0.8\GeV$). So in some sense, part of the hadronisation mass effects
will already be implicitly included in the parton showering and a
straightforward investigation of the difference between hadron and
parton levels will show up only a part of the mass effects (and there
will be a large ambiguity coming from the choice of scheme in which to
measure the parton level).

\section{Fits to data}
\label{sec:data}

\subsection{2-parameter fits}

\begin{figure}[tbp]
  \begin{center}
    \epsfig{file=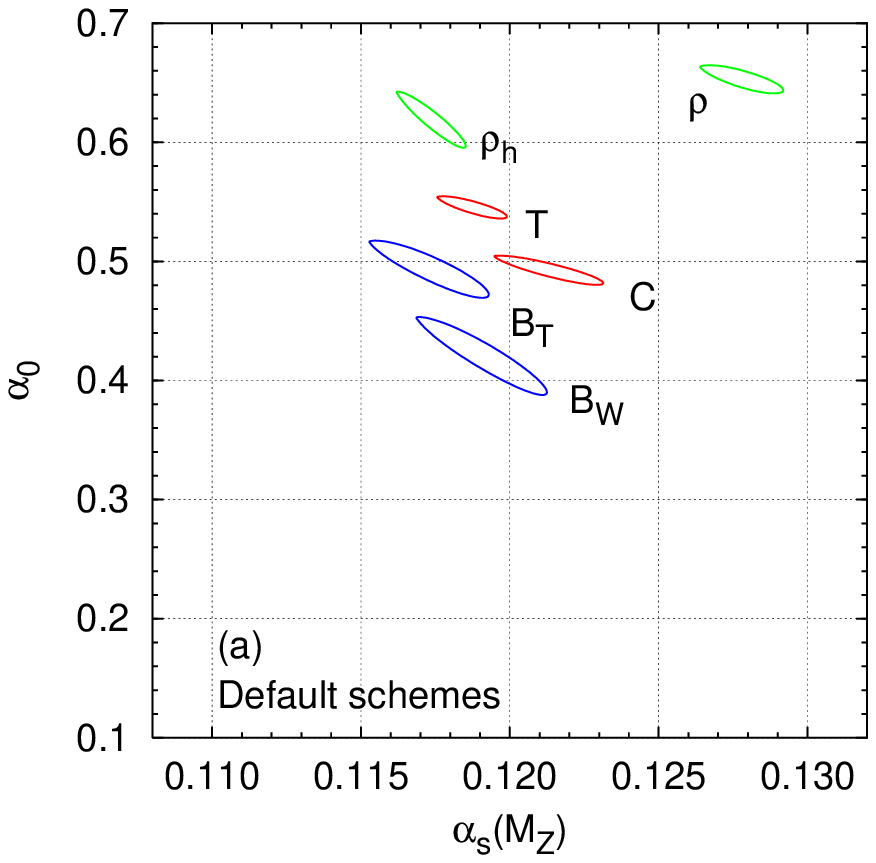,width=0.485\textwidth}\;
    \epsfig{file=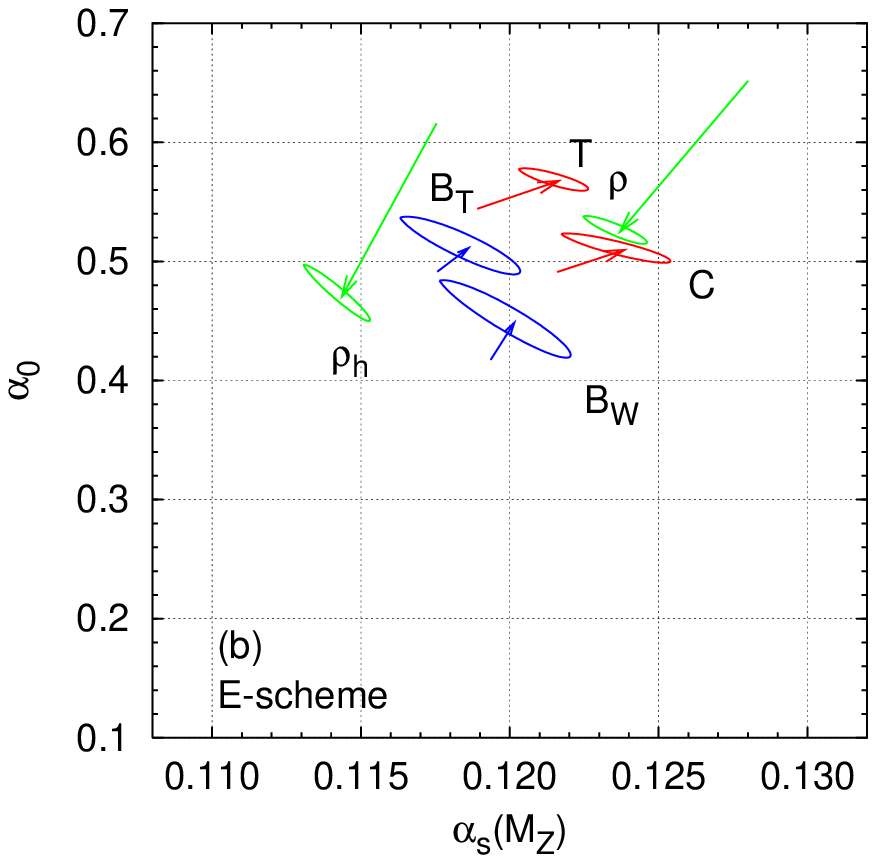,width=0.485\textwidth}
 \vspace{0.2cm} \\
    \epsfig{file=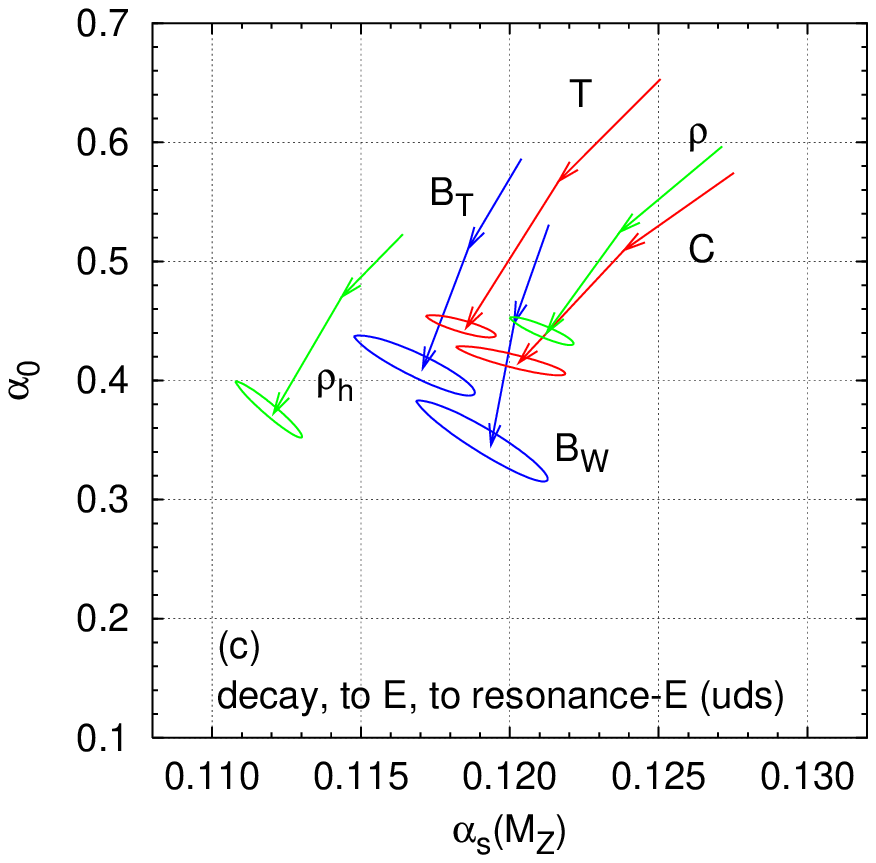,width=0.485\textwidth}\;
    \epsfig{file=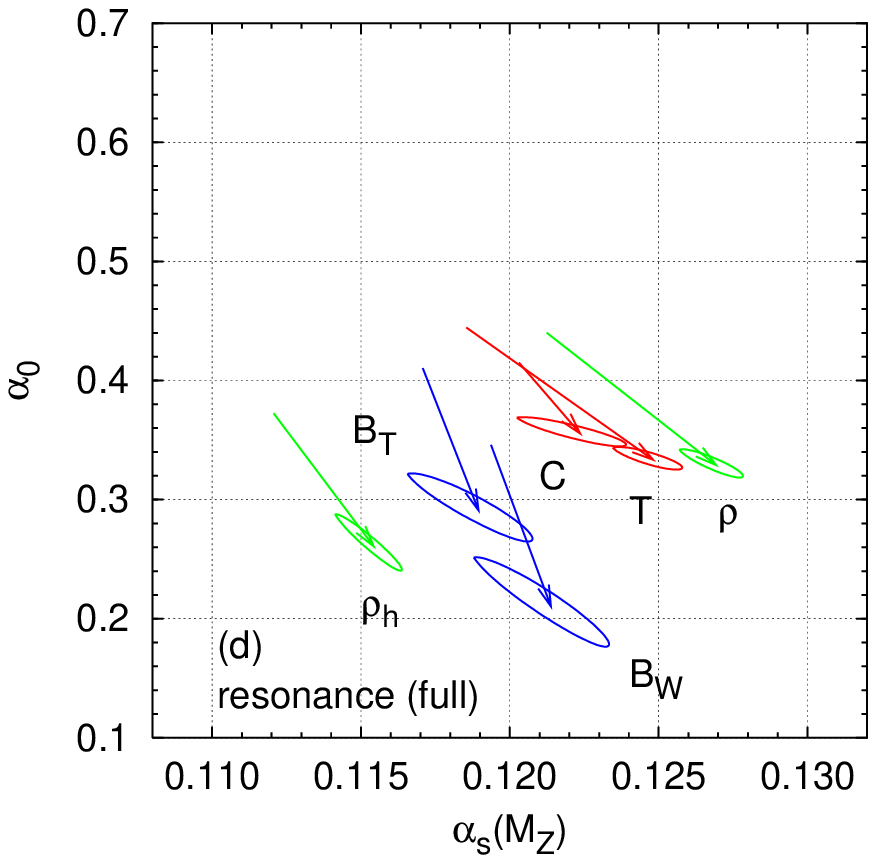,width=0.485\textwidth}
    \caption{1-$\sigma$ confidence-level contours from fits to
      event-shape variables in a range of schemes. (a) fits in the
      default schemes (normal hadron level); (b) fits in the
      $E$-scheme (normal hadron level), with arrows indicating the
      motion of the contour in going from the default to the
      $E$-scheme; (c) fits in the $E$-scheme at resonance level, with
      arrows indicating the motion of the contour from the
      decay-scheme, to the hadron-level $E$-scheme, to the resonance
      $E$-scheme --- here the correction to resonance level has
      carried out using only events with light primary quarks; (d)
      fits in the $E$-scheme at resonance level where the correction
      to resonance level now includes events with heavy primary quarks
      as well --- the arrows indicate the motion from the `uds'
      resonance level.}
    \label{fig:fitsall}
  \end{center}
\end{figure}

Following the suggestion of \cite{DW} it has become standard procedure
in recent years to carry out simultaneous fits for $\as$ and
$\alpha_0$, with formulae of the form
\begin{equation}
  \label{eq:basicfit}
  \langle \cV \rangle = C_1 \frac{\as}{2\pi} + C_2 \frac{\as^2}{(2\pi)^2} +
  c_\cV \frac{a_0}{Q} \,,
\end{equation}
where the $C_i$ are the perturbative coefficients for the mean value
and $a_0$ is defined in terms of $\alpha_0$ and $\as$ in
eq.~\eqref{eq:a0Toalpha0}. In order to set the 
scene we show in figure~\ref{fig:fitsall}a one-$\sigma$
confidence-level contours from such fits to data \cite{\alldata} for a
range of event-shape variables, all in the default schemes.  In the
absence of mass effects the universality hypothesis states that the
values of $\as$ and $\alpha_0$ should be consistent for the different
variables.

Compared to the figures of this kind that one usually sees, one
difference is the inclusion of a result for $\rho$ --- until now
generally only $\rho_h$ has been studied. While data do not exist for
$\rho$ itself, there are some data on the light-jet mass $\rho_l$, and
from this one can calculate $\rho = (\rho_h + \rho_l)/2$. What one
sees is a significant inconsistency between this variable and the
others.\footnote{It is perhaps ironic that this should be the one
  `standard' variable that had not been studied until now!}

Of course we know that we should really be carrying out the fits with
additional terms of the form \eqref{eq:abslnQterm}, so as to take into
account mass-dependent corrections. Let us for the time ignore
$\mu_\ell$ (\ie pretend it is conveniently zero!) and concentrate on
the term involving $\gamma_{\cV_S}\epsilon_\ell$ --- this piece is
measurement scheme-dependent. In the default schemes $\gamma_{\cV_S}$
is positive for the jet masses, and negative for all the other
variables (\cf table~\ref{tab:gammas}). If we ignore it then our fit
parameters for $\rho$ and $\rho_h$ should come out larger than for the
other variables. This is exactly what we see in
figure~\ref{fig:fitsall}a.

So if we want to be check universality we first have to ensure that
these non-universal mass effect are absent, \ie choose a scheme in
which the $\gamma_{\cV_S}$ are zero,
namely the $E$-scheme (we could also use a scheme in which all the
$\gamma_\cV$ are more or less proportional to $c_\cV$, such as the
$p$-scheme). Accordingly in figure~\ref{fig:fitsall}b we repeat the 
fits for $\as$ and $\alpha_0$ but using $E$-scheme data.\footnote{As
  discussed in section~\ref{sec:MCissues} there is as yet very little
  data in the $E$-scheme (in $e^+e^-$ it exists only from
  \textsc{Delphi}, for the jet-masses \cite{Delphi}), so we use
  Ariadne to correct data from the default scheme to the $E$-scheme.
  To be as close as possible to what the experiments use one might
  have preferred Pythia, but as we have seen in
  section~\ref{sec:MCcomparison} this does not reproduce the correct
  energy dependence for mass effects, though at phenomenologically
  relevant energies the discrepancy is quite small. All other corrections
  to different schemes and hadron levels are also done with Ariadne.
  It should be kept in mind that, especially at low values of $Q$
  there are big differences for example between Herwig and Ariadne ---
  consider $\tau \;(d-E)$ in figure~\ref{fig:HerwigvPythia} --- this
  implies non-negligible uncertainties regarding the effect of the
  scheme changes on the best-fit values for $\as$ and $\alpha_0$.} The
arrows show how the best fit values have moved in going from the
default to the $E$-schemes.

The switch to the $E$-scheme decreases the values of the jet masses,
while it increases, by somewhat less (in accord with the opposite sign
and smaller value of $\gamma_\cV/c_\cV$), the values of the other
variables. We see a change in both $\as$ and $\alpha_0$ because it is
only through a linear combination of the $\as$ and $\alpha_0$
$Q$-dependences that the fits can mimic a term of order $(\ln ^A
Q)/Q$. The `angle' of the arrows depends on the relative amounts of
$(\ln ^A Q)/Q$ and plain $1/Q$ in the mass-correction: if mass effects
involved just $1/Q$ corrections then only $\alpha_0$ would change. For
the broadenings the situation is more complex because the `universal'
power correction goes as $1/(\sqrt{\as} Q)$ which is more similar to a
mass effect than a pure $1/Q$ term, so there is less need for a change
in $\as$ to mimic the mass effect.

In the $E$-scheme the universality picture changes with respect to the
default schemes: whereas in the default scheme $\rho$ was clearly
inconsistent with the other variables, in the $E$-scheme it is now
very close to the thrust and the $C$-parameter. The heavy-jet mass on
the other hand now seems to be the least consistent of the different
variables as can be verified by examining the $\chi^2$
contribution from $\rho_h$ in a simultaneous fit to all variables. It
may well be that the non-inclusiveness of this variable is responsible
for its different behaviour, as is discussed in appendix~\ref{sec:mh}.

Now that we are in a scheme in which non-universal mass corrections
have been eliminated we can turn our attention to the question of
universal mass corrections, \ie the contribution related to $\mu_\ell$
in eq.~\eqref{eq:abslnQterm}. A first question of interest is whether
we actually need to worry about this term at all --- maybe it is
sufficiently small that it can be ignored altogether. We know that
$\mu_\ell$ depends on the hadronisation level $\ell$. To gauge the
importance of $\mu_\ell$  we study
three levels, each in the $E$-scheme: the decay level (actually
the decay scheme of the usual hadron level), the usual hadron level
and a `resonance' level. The latter is taken (arbitrarily) to consist
of the first level hadrons produced in the Pythia/Ariadne string
hadronisation routines. The results of 2-parameter fits to these
different hadronic levels are shown in figure~\ref{fig:fitsall}c: the
arrows start from the decay level best fit, go to the usual hadron
level best fit, and then to the resonance level, for which we also
show the $1$-$\sigma$ contours.

All variables move more or less in the same direction and by the same
amount --- this is consistent with our knowledge that $\mu_\ell$ is
multiplied by $c_\cV$ (the broadenings are more complex and move a bit
differently). Accordingly the situation regarding universality is
essentially unchanged (if anything, in the resonance level it is
somewhat improved). However 
in going from the decay to the resonance scheme $\as$ changes by
up to $0.007$; $\alpha_0$ is also sensitive to the hadronic level
chosen and varies by up to $0.2$. At $Q = M_Z$ the variation in the
observables themselves is of the order of $10$ to $15\%$. In other
words mass effects, even in a `universal' scheme, are responsible for
a significant part of an observable's value and have a non-trivial
effect on fits for $\as$ and $\alpha_0$.

\begin{figure}[tbp]
   \mbox{\!} \epsfig{file=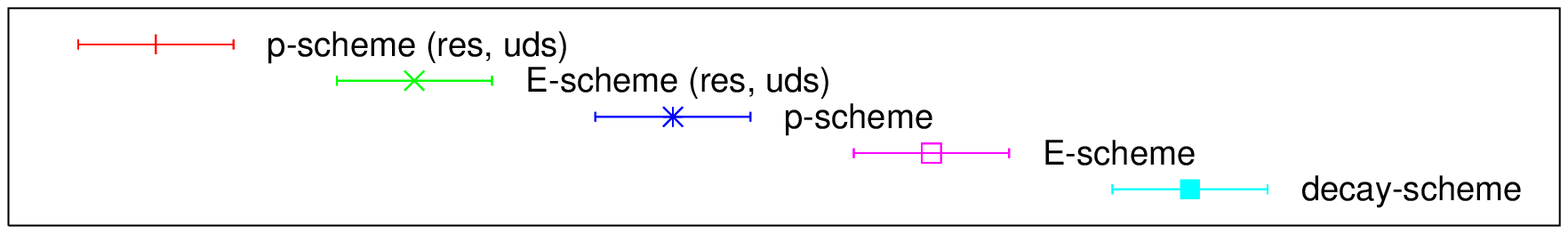,width=0.975\textwidth}
  \begin{center}
    \epsfig{file=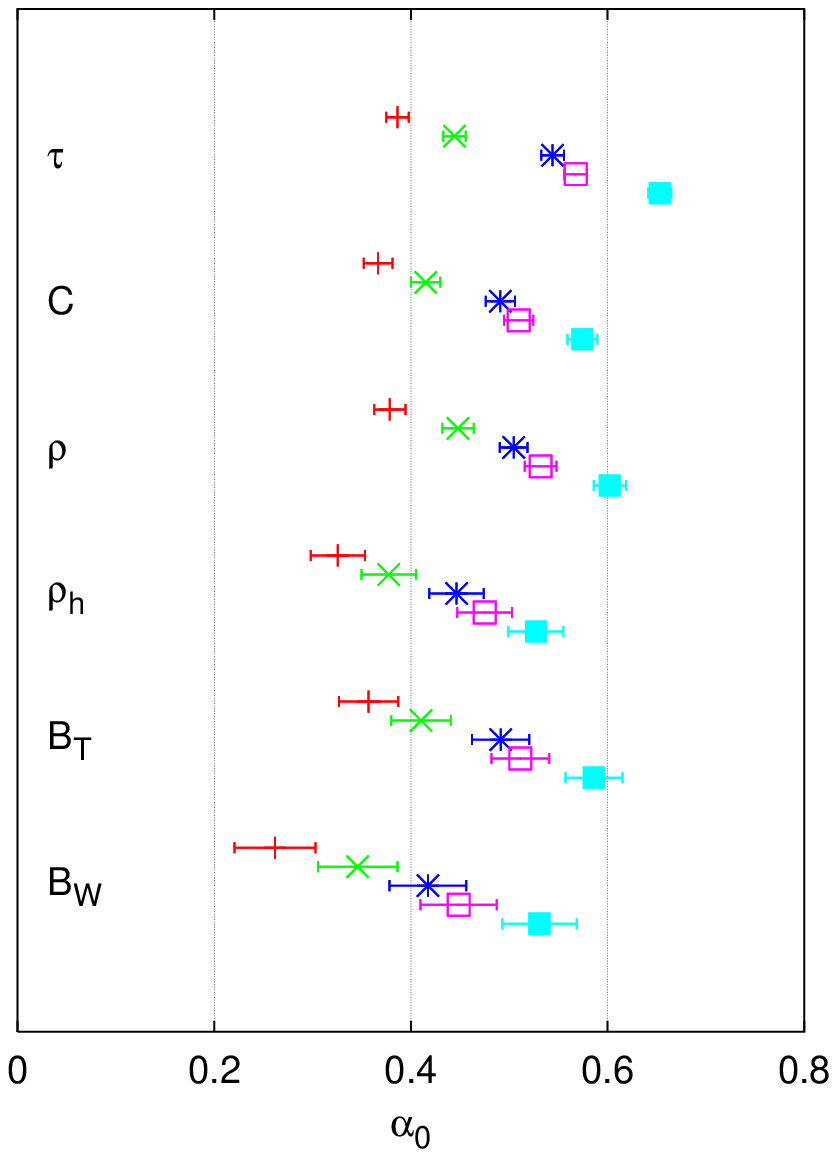,width=0.48\textwidth}\hfill
    \epsfig{file=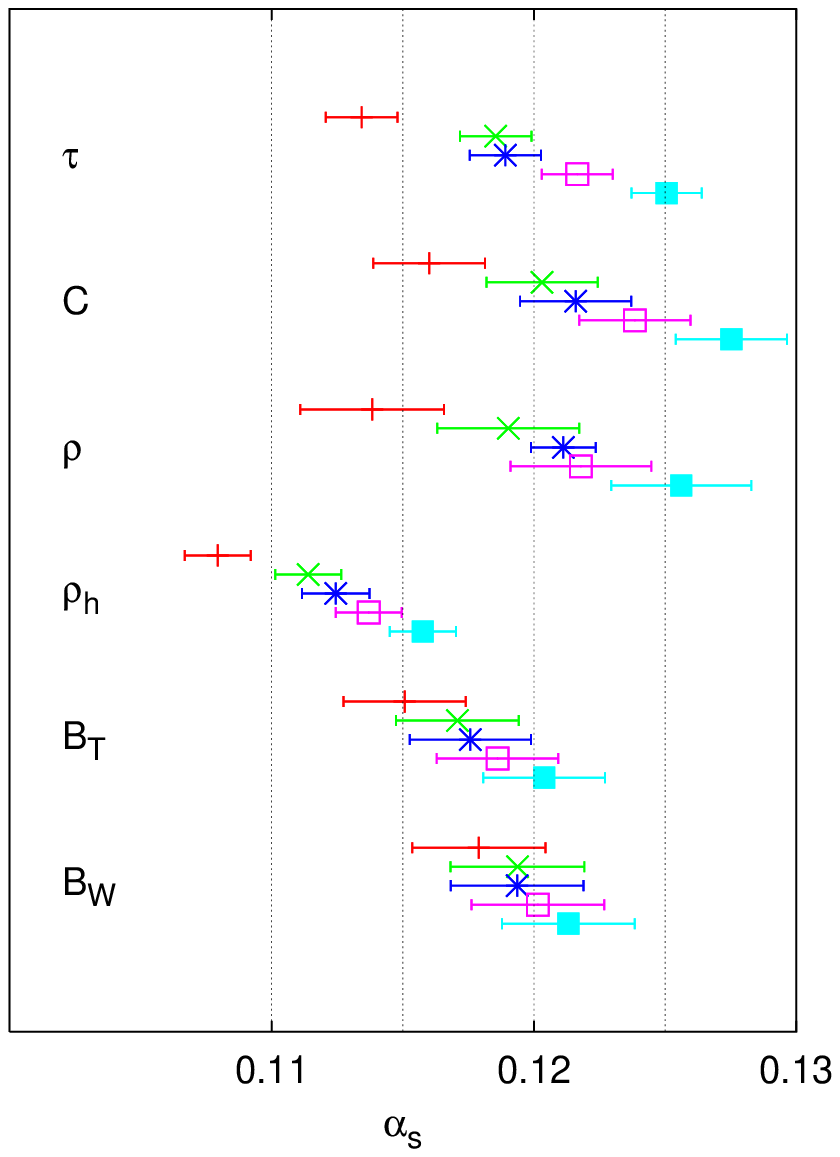,width=0.48\textwidth}
    \caption{Plots showing $\as$ and $\alpha_0$ values obtained by
      fitting to data corrected to a variety of schemes and
      hadronisation levels. Corrections relative to the default
      schemes have been obtained using Ariadne, as discussed in the
      text.}
    \label{fig:fitscatter}
  \end{center}
\end{figure}

If we had examined the same three hadronic levels in the $p$-scheme we
would have seen even larger dependence on the scheme because of the
contribution from the $\gamma_{\cV_{p}} \epsilon_\ell$ 
term (recall that
$\epsilon_\ell$, which is zero in the decay level, increases as
one goes towards the resonance level, \cf
eq.~\eqref{eq:muplusepsilon}). In particular the dependence of $\as$
on the hadronic level should double, a consequence of the relation
$\gamma_{\cV_{p}} \simeq - \gamma_{\cV_{\decay}}$. 
The complete range of
values for $\as$ and $\alpha_0$ in the different hadronic levels and
$E$ and $p$ schemes is summarised for our different event-shape
variables in figure~\ref{fig:fitscatter}.

Coming back to figure~\ref{fig:fitsall}c, one important point which we
have yet to mention is that the corrections from normal hadron to the
resonance level have been calculated using events with only light
(uds) primary quarks. The corrections have then been applied to all
events (including those with heavy primary quarks).  This is
equivalent to reconstructing all resonances except those associated
with primary heavy quarks.  The reason for doing this is that in an
event with heavy primary quarks, going to the resonance level involves
a reconstruction of the heavy-quark hadrons.  In the usual hadron
level these have decayed and the invariant mass $m_q$ of the hadron
contributes to the event shape at the level $m_q^2/Q^2$ (for a more
detailed discussion, see appendix~\ref{sec:heavyquarks}, and also the
discussion of heavy-quark decays in the context of fragmentation
functions in \cite{Nason:1994xx}), whereas in the resonance level ($E$
or $p$) schemes the mass of a forward moving hadron has little impact
on the value of the observable (other than at order $\as m_q^2/Q^2$).
If we had shown figure~\ref{fig:fitsall}c including heavy-quark events
to carry out the correction from hadron to resonance level, the
combination of the heavy-quark decay effect and the usual light-hadron
mass effects would have made it difficult to interpret the figure.
This is the reason why we applied a correction calculated using only
events with light primary quarks.

In order to give the `full story', in figure~\ref{fig:fitsall}d we
show the $1$-$\sigma$ contours which come from including primary heavy
quarks in our correction to resonance level (the arrows come from the
`uds' resonance level). The arrows are roughly at right angles to
those in the other plots --- this is closely linked to the fact the
difference between the `uds' and full resonance levels is a $1/Q^2$
correction (\ie something which dies off faster than $1/Q$, so that we
need the difference between a $1/Q$ and an $\as(Q)$ term in order to
approximate it over the given energy range). Despite their significant
effect on the best fit for $\as$, the heavy-quark decay changes the
value of the observables significantly only at low $Q$ values. At
$M_Z$ for example the effect on the thrust is $2\%$ (compared to
$16\%$ from the decay of all the light-resonances).

A final point worth mentioning is the following: if one believes, as
implied by Ariadne in figure~\ref{fig:totalhadr}, that of the various
hadronic levels the decay-level is that with the smallest
contamination from mass effects, then choosing this hadronic level for
a fit to $\as$ and $\alpha_0$, and accounting also for heavy-quark
decays (by either reconstructing the heavy-quark hadron, or explicitly
including the contribution from heavy-quark decay as calculated in
appendix~\ref{sec:heavyquarks}) will lead us to high results for
$\as$, around $0.130$.  This means that there may well be room for the
large higher-order perturbative coefficients predicted by Gardi and
Grunberg in \cite{GG}, which for normal $p$-scheme measurements imply
a value for $\as$ of about $0.110$.

\subsection{3-parameter fits}
\begin{table}[tb]
  \begin{center}
    \begin{tabular}{|l|c|c|c|c|}
      \hline
      $\cV$ & $\as$ & $\alpha_0$ & $\mu_\ell$ & $\chi^2/$d.o.f.\\ \hline
      $\tau$ res.\ (uds) &
       $ 0.1448\pm 0.0141$ & $ 0.891\pm0.237 $ & $-0.077\pm0.042$ 
        & $56.2 / 44$ \\ \hline
      $\tau$ res.\ (full) & 
       $0.1206 \pm 0.0095 $ & $0.262 \pm0.161 $ & $\nosign0.012 \pm 0.027$ 
       & $57.7 / 44$  \\ \hline
    \end{tabular}
    \caption{Results from a 3-parameter fit to the thrust in two
      different hadronic levels (`uds' and full resonance levels,
      $E$-scheme).}
    \label{tab:3parm}
  \end{center}
\end{table}

That fact that the event-shape values depend significantly on the
particular hadronic level chosen implies an important contribution to
the event-shape value from a term proportional to $\mu_\ell$. This
means that we should really be fitting for $\mu_\ell$ as well as for
$\as$ and $\alpha_0$, using an equation of the form
\begin{equation}
  \label{eq:basicfit3}
  \langle \cV \rangle = C_1 \frac{\as}{2\pi} + C_2 \frac{\as^2}{(2\pi)^2} +
  c_\cV \frac{a_0}{Q}  
  + c_\cV \frac{\mu_\ell}{Q} \ln^A \frac{Q}{\Lambda}\,.
\end{equation}
In the last term we fix $\Lambda=0.2\GeV$ (tying $\Lambda$ to the value of
$\as(M_Z)$, in practice the replacement of $\ln \frac{Q}{\Lambda}$ by
$2\pi/(\beta_0 \as)$, makes little difference). It turns out that
the inclusion of the last term makes the fitting procedure quite
unstable, and leads to large (correlated) errors on the individual fit
parameters as well as a strong dependence on subleading effects.
Consequently the significance of the results is limited.

To illustrate the point, we consider the thrust at two hadronic
levels (both in the $E$-scheme): the `uds' resonance level and the
`full' resonance level.  The former is subject to the $m_b^2/Q^2$
corrections arising from heavy quark decay, whereas the latter should
be free of them (but will still have other $1/Q^2$ corrections). The
fit results 
in the two cases, table~\ref{tab:3parm},
give completely different pictures. In
the `uds' resonance level, the fit results seem inconsistent with our
expectations for $\as$ (if only at $2$-$\sigma$). In the full
resonance level the value for $\as$ is `as we would like', but the
$\chi^2$ is a bit larger. 

There may of course be other subleading effects that we have not
yet considered (for example higher-order perturbative corrections)
which could cause further big changes. Additionally the event
generator used to 
correct to a given scheme may not have the right behaviour at low
values of $Q$. So the systematics are such that, at least currently,
it is difficult to extract reliable information from a $3$-parameter
fit: the uncertainties on $\as$ are even larger than those which arise
by considering a range of hadronic levels in a 2-parameter
$\as,\alpha_0$ fit and $\mu_\ell$ remains unconstrained.

If our main aim is to determine $\mu_\ell$ then we can try a
2-parameter fit with $\as$ fixed. But here again we find that the
systematic uncertainties on $\mu_\ell$, are of the same order as the
parameter expected size of the parameter itself, \ie about $0.02$ (\cf
the values of $\epsilon$ in figure~\ref{fig:mainresults}).

\section{Conclusions}
\label{sec:concl}

\paragraph{The Good.} %
In this paper we have understood many features of the contributions to
event-shapes that are associated with hadron masses: there are two
classes of contribution, both of which scale as $(\ln Q)^A/Q$. The
contribution from the `non-universal class' depends on the details of
whether the variable is defined in terms of $3$-momenta, energies and
angles, or a mixture of the two, and we can calculate the relative
magnitude of the mass-correction for different definition schemes. It
turns out that there is a privileged scheme ($E$-scheme) in which
non-universal mass corrections are absent, because their coefficient
is zero. The `universal' class of mass corrections gives contributions
proportional to the same coefficient $c_\cV$ that appears in
calculations of traditional universal (massless) $1/Q$ corrections;
universal mass corrections are present regardless of the definition of
the event-shape variable, and they are proportional to a new
non-perturbative parameter $\mu_\ell$ which depends on the particular
hadronic level $\ell$ (\ie stage of the decay chain) at which we
observe the event.

In traditional power correction analyses, $1/Q$ power corrections are
often given a quasi-perturbative interpretation in terms of an
infrared-finite coupling, reflecting the fact that they are associated
with the strictly perturbative concept of the renormalon divergence of
perturbation theory. The parameter $\alpha_0$ that appears in such
analyses can be related to a moment of the coupling in the infrared.
On the other hand our new parameter $\mu_\ell$ is more related to the
`reshuffling' of momenta associated with the production of mass --- in
this sense it is a much more intimately associated with the dynamics
of hadronisation, and there is perhaps even a possibility that its
determination could give us qualitative information about
hadronisation. It is not currently clear whether perturbative
contributions could also give contributions with a similar
$Q$-dependence.

Our analytical predictions for differences between different
measurement schemes agree well with results from two commonly used
Monte Carlo event generators, Herwig and Ariadne. Furthermore, in fits
for $\as$ and $\alpha_0$ from a range of event-shape variables, they
explain a pattern observed both in $e^+e^-$ and DIS of the jet masses
giving significantly larger values of both $\as$ and $\alpha_0$ than
other variables: this is because in the default schemes there are
positive (non-universal) mass-corrections for the jet masses and
negative corrections for the other variables. Measuring all variables
in the $E$-scheme leads to a significant improvement in the
consistency between the jet masses and the other variables. We also
note that for the jet masses the $p$ and $E$-schemes are relatively
insensitive to certain experimental systematics (associated with
difficulties in identifying hadrons) which are relevant in the default
measurement scheme.

\paragraph{The Bad.} %
Unfortunately it seems that a direct experimental verification of the
properties of mass effects is quite difficult. In principle mass
effects can be seen by looking at the difference between two
measurement schemes for the same variable (say massive and $E$-schemes
in the case of the jet mass). However the experimental determination
of such differences relies on the accurate identification of the mass
of each particle in an event, whereas the experimental procedure
usually just involves the assignment of the pion mass to each hadron.
Since a large part of mass effects seems to come from kaons and
nucleons this is a poor approximation. Given that the results are then
corrected for what cannot be seen, using Monte Carlo programs, the
resulting `measurement' of the difference between two schemes is
likely to be as much a reflection of the properties of the Monte Carlo
events as of the actual events!\footnote{One possible more direct
  experimental test of the ideas at the base of the predictions for
  mass effects would be a measurement of a quantity such as the sum of
  the inverse energies of all nucleons (or some other particle
  species).  This should scale as $(\ln Q)^A$.}

It so happens that in one of the Monte Carlo programs most commonly
used for correcting data, Pythia, the $Q$-dependence of mass effects
differs significantly from our predictions at very high energies,
going very roughly as $(\ln \ln Q)/Q$ rather than $(\ln Q)^A/Q$.  This
is probably related to approximations in Pythia's implementation of
coherence.  Fortunately for measurements in the $p$ or $E$-schemes
this should not have too large an effect on the practical
determination of event shape values since the 3-momentum or energy are
close to what is truly measured by detectors (it is the difference
between them that is poorly measured). Furthermore at today's energies
the discrepancy in the Pythia energy dependence is a small effect
compared to other differences between event generators.

In general, given the difficulties of a direct measurement of mass
effects in event shapes, for an experimental test of the picture
outlined here it might be worth investigating the feasibility of
measuring some other observable expected to depend on the same
$(-1)^\mathrm{th}$ moment of particle energies. The simplest might be
$\sum_i 1/E_i$ where the sum runs over charged tracks. This might give
an idea of whether the LPHD hypothesis used in this paper works also
for negative moments, where it has so far never been tested.

One `negative' implication of our results relates to the determination
of the perturbative and non-perturbative parameters of QCD. We would
like to measure our new non-perturbative parameter $\mu_\ell$. But the
degeneracy in a 3-parameter fit for $\as$, $\alpha_0$ and $\mu_\ell$
is such that the currently available data are not precise enough to
give us any meaningful constraints on any of the fit parameters.
Furthermore the fit results are very unstable with respect to
systematic uncertainties.  Given this limitation we might decide that
for the time being we should carry on as before with two-parameter
fits $\as$ and $\alpha_0$ (of course with the event-shape variables
now measured in the $E$-scheme).  Our having neglected a parameter in
the fit will translate to systematic errors on $\as$ and $\alpha_0$.
We can try to gauge the size of the systematics by carrying out the
fit on results at different levels of the hadronic decay chain (using
Monte Carlo results to determine the correction) --- this suggests
systematics of the order of $\pm0.004$ on $\as$ and of about $\pm 0.1$
on $\alpha_0$.

\paragraph{The Ugly.} In the course of our studies we have come
across two effects, unrelated to light-hadron masses, that have a
significant impact on event-shape studies and so warrant further
investigation.

Firstly there is the observation that heavy-quark effects have a large
impact on fit values for $\as$ and $\alpha_0$ --- measuring an event
shape before and after heavy quark decay modifies $\as$ values by up
to $0.006$, even though at $M_Z$ the effect of heavy quark decay is
only $2\%$. This peculiar behaviour is seen because the $m_q^2/Q^2$
contribution from quark decay
is simulated by an increase of $\as$ and a decrease of $\alpha_0$.
Given that the tools for studying heavy-quark effects in event shapes
are well-established, there is a strong argument for carrying out
analyses that make full use of them.


Secondly fits for the heavy-jet mass (a very non-inclusive variable)
lead to values for $\as$ which are about $10\%$ smaller than for
inclusive variables like the thrust or the mean jet mass. This needs
to be understood. It could be due to a difference in the behaviour of
the perturbation series at higher orders. But in appendix~\ref{sec:mh}
there is evidence from Monte Carlo simulations that hadronisation
corrections for $\rho_h$ have unusual characteristics: in contrast to
what is seen in more inclusive variables, the hadronisation depends
strongly on the underlying hard configuration. There is therefore a
need to develop techniques allowing a more formal approach to the
study of such problems.

\section*{Acknowledgements}

We would like to thank Hasko Stenzel and Bryan Webber for discussions
about differences between the jet masses and the thrust which where
the original motivation for this work. We are also grateful to them
and additionally to Mrinal Dasgupta, Yuri Dokshitzer, Klaus Hamacher,
Einan Gardi, Pino Marchesini, Klaus Rabbertz, Torbj\"orn Sj\"ostrand
and Giulia Zanderighi for valuable discussions and suggestions
throughout the course this work. One of us (GPS) would further like to
thank the Milano section of the INFN and Milano and Milano-Bicocca
universities for the use of computing facilities and hospitality while
part of this work was carried out.

\appendix

\section{Summary of notation}
\label{sec:notation}

For convenience we give here a summary of the definition of the various
schemes introduced in this article.
\bigskip

\newcommand{\twomini}[2]{
  \begin{minipage}[t]{0.2\textwidth}
    #1
  \end{minipage}
  \begin{minipage}[t]{0.75\textwidth}
    #2
  \end{minipage}\vspace{0.4cm}
}

\twomini{$\boldsymbol{p}$\textbf{-scheme}}{Scheme in which the
  observable is defined solely in terms of particle 3-momenta.}

\twomini{$\boldsymbol{E}$\textbf{-scheme}}{Scheme in which the
  observable is defined solely in terms of particle energies and
  angles.}

\twomini{\textbf{decay-scheme}}{Scheme in which all massive particles
  are decayed isotropically into pairs of massless particles. The
  observable is then calculated using the resulting ensemble of
  massless particles.}

\noindent We also summarise 
some of the other notation used and introduced in this
article.\bigskip

\twomini{$\boldsymbol{\cV}$}{An event-shape variable.}

\twomini{$\boldsymbol{\cV_{p}, \cV_{E}, \cV_{\decay}}$}{An event-shape
  variable in $p$, $E$ or decay-scheme, respectively.}

\twomini{$\boldsymbol{c_\cV}$}{The coefficient of the `traditional'
power correction for the observable $\cV$, introduced in
eq.~\eqref{eq:cV} and given for a range of observables in
table~\ref{tab:cvs}.}

\twomini{$\boldsymbol{\langle \delta_m \cV \rangle}$}{The 
non-universal mass-dependent correction to the mean value of the
observable $\cV$, \cf eq.~\eqref{eq:NPmassiveGen}.}

\twomini{$\boldsymbol{\delta c_\cV(m^2/k_t^2)}$}{The modification to
$c_\cV$ for a particle with a given $m^2/k_t^2$, \cf
eq.~\eqref{eq:deltacVmassive}.}

\twomini{$\boldsymbol{\gamma_\cV}$}{The coefficient of the leading
$m^2/k_t^2$ dependence of $\delta c_\cV(m^2/k_t^2)$ for $k_t \gg m$,
as defined in eq.~\eqref{eq:GammaExpnd}. It is shown for a range of
variables in their default schemes in table~\ref{tab:gammas}.}

\twomini{$\boldsymbol{X_\decay(m^2/k_t^2)}$}{The mean relative change in
the sum of $|k_t|$'s due to the decay of a massive particle of mass
$m^2$ and transverse momentum $k_t^2$, as defined in
eq.~\eqref{eq:genXdef}.}

\twomini{$\boldsymbol{A}$}{Shorthand for
$4\CA/\beta_0$. Mass-dependent corrections have a leading term
proportional to $(\ln Q/\Lambda)^A/Q$.}

\twomini{$\boldsymbol{\epsilon_\ell}$}{The normalisation of the
non-universal mass-dependent correction for a hadronic level $\ell$,
\cf eqs.~(\ref{eq:epsilon}, \ref{eq:abslnQterm}).}

\twomini{$\boldsymbol{\mu_\ell}$}{The normalisation of the
universal mass-dependent correction for a hadronic level $\ell$, \cf
eqs.~(\ref{eq:abslnQterm}, \ref{eq:defmuell}).}


\section{Broadenings in the decay scheme}
\label{sec:broad}

In determining the QCD-based predictions for mass-effects in
section~\ref{sec:QCDappl} we made the assumption that for all
variables considered, mass effects matter only in the region of small
rapidities. As a result we could ignore the rapidity dependence of the
hadron distribution, because it is significant only at large
rapidities. More precisely we expanded the rapidity dependence of
$\Phi_{h,-\om}(\eta,Y)$ and showed that it gave terms suppressed by a
power of $\as$,
\begin{equation}
  \label{eq:whyconverge}
  \int d\eta  \; \delta f_{\cV,\om}(\eta)\; \as \eta = \cO{\as},
\end{equation}
which holds if $ \delta f_{\cV,\om}(\eta)$ decreases sufficiently
rapidly with $\eta$.  This works for all variables except the
broadenings in the decay scheme, the reason being that for the
broadenings the difference between the normal and decay schemes is
sensitive to particle masses for particles up to rapidities of order
$1/\sqrt{\as}$.  For the example the total jet broadening has
\begin{equation}
  \label{eq:dfBT}
  \delta f_{B_{T,\mathrm{decay}},\om}(\eta) =  f_{B_T}(\eta)\, X_{\decay,\om}
\end{equation}
where we can write \cite{Broad}
\begin{equation}
  \label{eq:fBT}
   f_{B_T}(\eta) \simeq \int_0^\infty d\ell\, \Theta\left(\ell +\eta_0 -
     \eta \right) \Theta(\eta)  \;
      \left( \frac{2\as\CF \ell }{\pi} 
        + \frac{\CF\as^2 \beta_0 \ell^2}{(\pi)^2}
      \right)
              e^{ -\frac{\as\CF\ell^2 }{\pi} - \frac{\CF\beta_0 \as^2
                  \ell^3}{3\pi^2}}\,. 
\end{equation}
(We recall that here $\eta_0$ has been defined with an additional
$3/4$ compared to what is given in \cite{Broad}).  Since $f_{B_T}$ is
roughly $1$ up to $\eta \sim 1/\sqrt{\as}$ the integral
\eqref{eq:whyconverge} is of order $1$ (while the leading term is of
order $1/\sqrt{\as}$, \cf table~\ref{tab:cvs}). Even though this term
is subleading, in analogy with what is done for the $c_\cV$'s, we wish
to control it, which can be achieved as follows.  First we write
\begin{equation}
 \int d\eta\,
    \,\delta f_{\cV,\om} (\eta)\,
    \Phi_{h,-\om}(\eta,Y) = 
\Phi_{h,-\om}(0,Y) \delta c_{\cV,\om}
\end{equation}
where the $\eta$ dependence of $\Phi_{h,-\om}$ has been absorbed into
$\delta c_{\cV,\om}$: 
\begin{equation}
  \delta c_{\cV,\om} = \int d\eta \; \delta f_{\cV_{\mathrm{decay}},\om}(\eta)
       \left(1 - \frac{A}{\om}\frac{\eta}{Y}  + \cdots \right)\,.
\end{equation}
Substituting in the above equations for $\delta f_{B_{T,\mathrm{decay}},\om}$
gives
\begin{equation}
  \label{eq:cbtdecay}
  \delta c_{B_{T,\mathrm{decay}},\om} =  \left(
    \frac{\pi}{2\sqrt{\CF \as}} -\frac{\beta_0}{6\CF} + \eta_0 
    - \frac{A}{\om}\frac{\beta_0}{4\CF} \right) X_{\decay,\om}\,.
\end{equation}
In section~\ref{sec:decayscheme} we had stated that
\begin{equation*}
  \delta  c_{\cV_{\mathrm{decay}},\om} =  c_\cV \, X_{\decay,\om}\,.
\end{equation*}
If we compare our result \eqref{eq:cbtdecay} with $c_{B_T}$ in
table~\ref{tab:cvs}, we find that there is an additional term $
-\frac{A\beta_0}{4\CF\om} = -\frac{\CA}{\CF\om}$, not present in
$c_\cV$. This is the piece which arises from the rapidity dependence
of the hadron distribution and in practice it gives quite a large
correction to $ \delta c_{B_{T,\mathrm{decay}},\om}$.

The corresponding derivation for the wide-jet broadening involves the
use of
\begin{equation}
  \label{eq:dfBW}
  \delta f_{B_{W,\mathrm{decay}},\om}(\eta) =  f_{B_W}(\eta)\, X_{\decay,\om}\,,
\end{equation}
where 
\begin{equation}
  \label{eq:fBW}
   f_{B_W}(\eta) \simeq 
       \frac12 \int_0^\infty d\ell\, \Theta\left(\ell +\eta_0 -
     \eta \right) \Theta(\eta)  \;
      \left( \frac{4\as\CF \ell }{\pi} 
        + \frac{2\CF\as^2 \beta_0 \ell^2}{(\pi)^2}
      \right)
              e^{ -\frac{2\as\CF\ell^2 }{\pi} - \frac{2\CF\beta_0 \as^2
                  \ell^3}{3\pi^2}}\,. 
\end{equation}
We thus obtain
\begin{equation}
  \label{eq:cbwdecay}
  \delta c_{B_{W,\mathrm{decay}},\om} =  X_{\decay,\om} \left(
    \frac{\pi}{4\sqrt{2\CF \as}} -\frac{\beta_0}{24\CF} + \frac{\eta_0 }{2}
    - \frac{A}{\om}\frac{\beta_0}{16\CF} \right)\,. 
\end{equation}

The final form for the broadening mass-dependent power correction is
analogous to eq.~\eqref{eq:Final} but with the appropriate
$\om$-dependence introduced for $\gamma_\cV$:
\begin{equation}
  \label{eq:FinalBroad}
  \langle \delta_{m} B_\decay \rangle = \gamma_{B_{\decay},A}\,
  \frac{\epsilon}{Q} \,\ln^A
  \frac{Q}{\Lambda}
    + \cdots
\end{equation}
with
\begin{equation}
  \gamma_{B_{T,\mathrm{decay}},\om} =  \frac14 \left(
    \frac{\pi}{2\sqrt{\CF \as}} -\frac{\beta_0}{6\CF} + \eta_0 
    - \frac{A}{\om}\frac{\beta_0}{4\CF} \right)\,.
\end{equation}
and 
\begin{equation}
  \gamma_{B_{W,\mathrm{decay}},\om} =  \frac14 \left(
    \frac{\pi}{4\sqrt{2\CF \as}} -\frac{\beta_0}{24\CF} + \frac{\eta_0 }{2}
    - \frac{A}{\om}\frac{\beta_0}{16\CF} \right)\,.
\end{equation}

\section{Heavy quark decay}
\label{sec:heavyquarks}

While effects due to heavy quarks are not strictly speaking the
concern of this article, it turns out that they can have a significant
effect on the fit results for $\as$ and $\alpha_0$. The presence of
heavy quarks affects both the perturbative and non-perturbative
contributions to the event shapes.

Perturbative calculations of event shapes involving heavy quarks have
been in existence for a few years now \cite{Nason:1998nw,
  Oleari:1997az, Rodrigo:1997ha, Rodrigo:1996gw, Rodrigo:1997gy,
  Bernreuther:1997jn, Brandenburg:1998pu}. They have started to be
used for experimental studies of events with heavy quarks, with
hadronisation corrections deduced from Monte Carlo event generators
(see for example \cite{Abbiendi:1999fs}).

Power corrections to event shapes with heavy quarks have been studied
in \cite{Zoltan} --- they differ from the light quark case because
very collinear radiation ($\eta \gtrsim \ln Q/m_q$, where $m_q$ is the
heavy-quark mass) is suppressed. For event shapes like $\tau$, $C$ and
the jet masses this leads to a reduction of the power correction by an
amount of order $ m_q \Lambda/ Q^2$. The broadenings are more complex.

For typical measurements of event shapes, in the presence of heavy
quarks there is a second `hadronisation' contribution due to the fact
that what is measured is not the final state involving charm or
bottom-quark hadrons, but rather a final state where the heavy-quark
hadrons have decayed. This has been discussed in some detail in the
context of fragmentation functions in \cite{Nason:1994xx}, and many
aspects turn out to be quite similar for event shapes. Since (for
$Q\gg m_q$) the quark is fast-moving, the effect of the decay is to
produce a bunch of nearly collinear hadrons whose invariant mass is
the heavy-hadron mass (for simplicity, from here on we neglect the
distinction between the heavy-quark and the heavy-quark hadron, while
bearing in mind that the $m_q$ relevant for the decay is actually the
hadron mass).

If we consider a Born configuration consisting of two back-to-back
heavy quarks then it is quite straightforward to see what effect the
decay will have on the simpler event shapes. In the massive scheme
$\rho$ is unchanged by quark decay. In the $p$ or $E$-schemes it goes
from being zero before the quark decay to
\begin{equation}
  \rho = \frac{m_q^2}{Q^2} 
\end{equation}
after quark decay. This result can also be used to deduce the
correction to the thrust and $C$-parameter because in the collinear
limit we have \cite{Catani:1998sf}
\begin{equation}
  \tau \simeq \rho_L + \rho_R \qquad C\simeq 6(\rho_L + \rho_R)\,.
\end{equation}
where $\rho_L$ and $\rho_R$ are the left and right-hemisphere jet
masses respectively. To summarise, the corrections expected as a
result of the decay of primary heavy quarks are shown in
table~\ref{tab:heavyquark}.

\begin{table}[tb]
\begin{center}
\begin{tabular}{|l|c|c|c|c|}\hline
$\cV$ & $\tau$ & $\rho$ & $\rho_h$ & C\\ \hline
$\delta \cV$ (from quark decay)  & 
  $\frac{2m_q^2}{Q^2}$ & $\frac{m_q^2}{Q^2}$ & $\frac{m_q^2}{Q^2}$ &
  $\frac{12m_q^2}{Q^2}$ \\ \hline
\end{tabular}
\end{center}
\caption{Corrections to event shapes arising from heavy quark decay.}
\label{tab:heavyquark}
\end{table}

For the jet broadenings the situation is more complex.  If we start
from the Born configuration and let the heavy quarks decay, then with
respect to the thrust axis all the decay products will have transverse
momenta of order $m_q$, leading to values for the broadenings of order
$m_q/Q$. However to consider just the effect of quark decay on the
Born configurations is an oversimplification. The variables discussed
above had the property that they are linear for soft and/or collinear
particles (even $\rho_h$, to within the approximations required here):
namely the effect of quark decay is the same regardless of whether we
have the Born configuration, or one with extra soft and/or collinear
particles.

But the broadenings do not have this property. What goes on with
heavy-quark decay is very similar to the dynamics that led to the
rather complex form for the power correction to the broadening.
Essentially the extra transverse momentum from the decay only
contributes at order $m_q/Q$ if the quark's angle with respect to the
thrust axis is less than $m_q /Q$. If the quark's angle is larger than
this, then azimuthal averaging causes the $m_q/Q$ correction to be
reduced to a $m_q^2/p_t Q$ correction, where $p_t$ is the quark's
transverse momentum. A more quantitative understanding would require a
full treatment of the 3-body heavy hadron decay and a study (from a
perturbative resummation) of how the quark transverse momentum
compares with $m_q$ as a function of $Q$ (for very large $Q$ it will
usually be much bigger, but phenomenologically accessible values of
$Q$ may not be large enough).

In \cite{Nason:1994xx} it has been pointed out that for the
longitudinal fragmentation functions there can also be corrections
proportional to $m_b/Q$ associated with the decay of secondary
heavy-quarks produced from a soft gluon, though it is suggested that
for today's energies such a behaviour may not yet have set in. The
possibility of a similar contribution in event shapes should be
investigated.

\section{The heavy-jet mass}
\label{sec:mh}

We observed in section~\ref{sec:data} that even in a `proper' scheme
the $(\as,\alpha_0)$ fits for the heavy-jet mass (and perhaps also the
wide-jet broadening) seem to some extent inconsistent with the
results for the other variables. The distinguishing feature of the
heavy-jet mass is its non-inclusiveness, since it measures a specific
hemisphere of the event (the heavy one), whereas other variables
measure the properties of the whole event.

We may well ask why non-inclusiveness leads to differences. One
interesting analysis has been presented in \cite{Korchemsky}, which
suggests that hadronisation corrections can be different in the two
hemispheres and convert a perturbatively light jet into a heavy one.
However this effectively \emph{increases} the power correction rather
than decreasing it and so cannot explain the relatively small $\as$
and $\alpha_0$ values that are observed. This does
not mean that such a mechanism is not present at all --- indeed in the
difference between the $E$ and $p$ schemes the heavy jet mass
correction is larger than that for the single jet mass (\cf
fig.~\ref{fig:mainresults}), and this could be due to such a mechanism
(it could also simply be because there are more hadrons in the heavy
hemisphere).

\begin{figure}[tbp]
  \begin{center}
    \epsfig{file=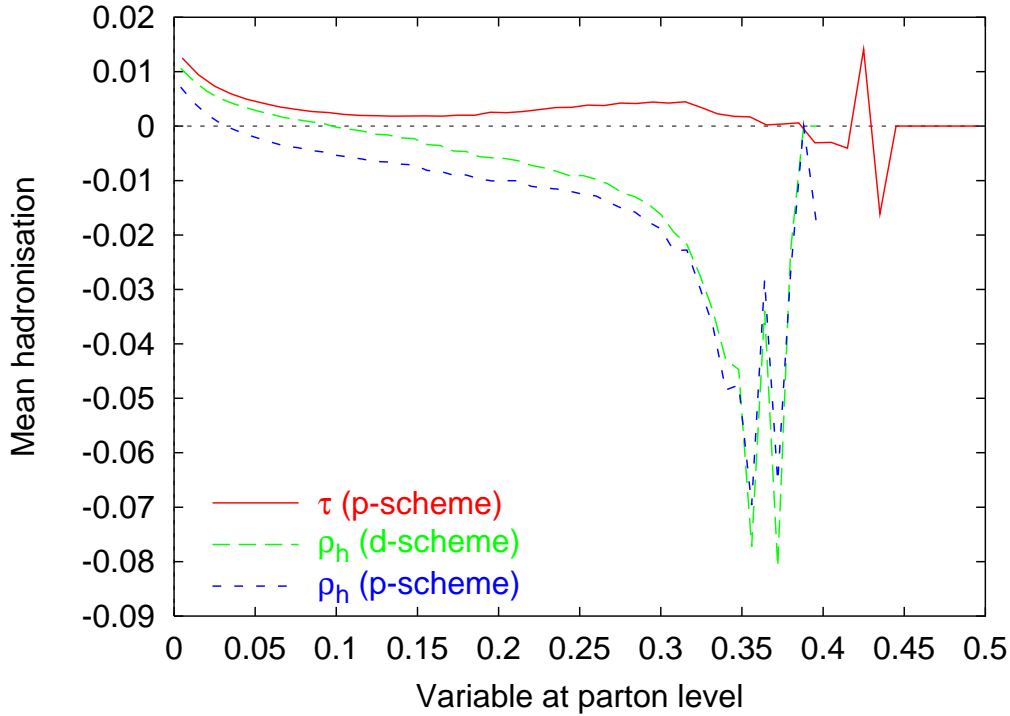,width=0.8\textwidth}
    \caption{The hadronisation correction as a function of the
      value of the variable at parton level. }
    \label{fig:hadr}
  \end{center}
\end{figure}

To help understand what is happening we have used Pythia to look at
the mean hadronisation as a function of the value of the variable at
parton level, figure~\ref{fig:hadr}. For the thrust, the hadronisation
is fairly independent of the parton-level thrust value. For the heavy
jet masses (in the massive and $p$-schemes) however the hadronisation
correction is very negative for larger values of $\rho_h$ at parton
level (we note that there are also very large event-by-event
fluctuations in the hadronisation --- this means that the
hadronisation does not just cause a simple shift of the perturbative
distribution). This feature may be at the root of the non-universality
seen in the heavy jet mass, and needs to be understood --- one
explanation might be that when one hemisphere is perturbatively heavy,
the phase-space that remains for further emissions is limited, and the
only way of emitting non-perturbative radiation in the
light-hemisphere is if some energy (and mass) is removed from the
heavy hemisphere. However such a hypothesis would need to be placed on
a more rigorous mathematical footing for it to be tested.

Such a phenomenon might also play a role in the wide-jet broadening,
where it has been observed that the theoretically predicted
distribution is too wide at larger values of $B_W$ \cite{SZ}.




\begin{thebibliography}{99}


\bibitem{Webber94}
B.~R.~Webber,
Phys.\ Lett.\  {\bf B339} (1994) 148
[hep-ph/9408222].

\bibitem{Webber94tube}
B.~R.~Webber,
in proceedings of the {\it Summer School on Hadronic Aspects of
  Collider Physics, Zuoz, Switzerland, 1994}, 
hep-ph/9411384.


\bibitem{DW}
Y.~L.~Dokshitzer and B.~R.~Webber,
Phys.\ Lett.\  {\bf B352} (1995) 451
[hep-ph/9504219].


\bibitem{DMW}
Y.~L.~Dokshitzer, G.~Marchesini and B.~R.~Webber,
Nucl.\ Phys.\  {\bf B469} (1996) 93
[hep-ph/9512336].

\bibitem{DasW}
M.~Dasgupta and B.~R.~Webber,
Phys.\ Lett.\  {\bf B382} (1996) 273
[hep-ph/9604388].

\bibitem{KS}
G.~P.~Korchemsky and G.~Sterman,
Nucl.\ Phys.\  {\bf B437} (1995) 415
[hep-ph/9411211];\\
G.~P.~Korchemsky, G.~Oderda and G.~Sterman,
presented at \emph{5th International Workshop on Deep Inelastic Scattering
and QCD (DIS 97)}, Chicago, IL, April 1997
hep-ph/9708346;\\
G.~P.~Korchemsky and G.~Sterman,
Nucl.\ Phys.\ B {\bf 555} (1999) 335
[hep-ph/9902341].

\bibitem{AZ}
R.~Akhoury and V.~I.~Zakharov,
Phys.\ Lett.\  {\bf B357} (1995) 646
[hep-ph/9504248].
\\
R.~Akhoury and V.~I.~Zakharov,
Nucl.\ Phys.\  {\bf B465} (1996) 295
[hep-ph/9507253].


\bibitem{BB}
M.~Beneke and V.~M.~Braun,
Nucl.\ Phys.\  {\bf B454} (1995) 253
[hep-ph/9506452].

\bibitem{NS} P. Nason and M.~H.~Seymour,
Nucl.\ Phys.\  {\bf B454} (1995) 291
[hep-ph/9506317].


\bibitem{BBM}
M.~Beneke, V.~M.~Braun and L.~Magnea,
Nucl.\ Phys.\  {\bf B497} (1997) 297
[hep-ph/9701309].

\bibitem{Milan}
Y.~L.~Dokshitzer, A.~Lucenti, G.~Marchesini and G.~P.~Salam,
Nucl.\ Phys.\  {\bf B511} (1998) 396
[hep-ph/9707532], erratum {\it ibid.} {\bf B593} (2001) 729.

\bibitem{Milan2}
Y.~L.~Dokshitzer, A.~Lucenti, G.~Marchesini and G.~P.~Salam,
JHEP {\bf 9805} (1998) 003
[hep-ph/9802381].

\bibitem{DasWMilan}
M.~Dasgupta and B.~R.~Webber,
JHEP {\bf 9810} (1998) 001
[hep-ph/9809247].

\bibitem{DMSmye}
M.~Dasgupta, L.~Magnea and G.~Smye,
JHEP {\bf 9911} (1999) 025
[hep-ph/9911316];\\
G.~Smye,
hep-ph/0101323.

\bibitem{Broad}
Y.~L.~Dokshitzer, G.~Marchesini and G.~P.~Salam,
Eur.\ Phys.\ J.\ direct {\bf C3} (1999) 1
[hep-ph/9812487], erratum {\it ibid.} {\bf C1} (2001) 1.


\bibitem{GG} E. Gardi and G. Grunberg,
JHEP {\bf 9911} (1999) 016
[hep-ph/9908458].


\bibitem{Zoltan}
Z.~Trocsanyi,
JHEP {\bf 0001} (2000) 014
[hep-ph/9911353].

\bibitem{Korchemsky}
G.~P.~Korchemsky and S.~Tafat,
JHEP {\bf 0010} (2000) 010
[hep-ph/0007005].



\bibitem{exppoweree} 
P.~Abreu {\it et al.}  (DELPHI Collaboration),
Phys.\ Lett.\ {\bf B456} (1999) 322;\\
P.~A.~Movilla Fernandez, O.~Biebel, S.~Bethke, S.~Kluth and P.~Pfeifenschneider
                  (JADE Collaboration),
Eur.\ Phys.\ J.\ {\bf C1} (1998) 461
[hep-ex/9708034];\\
M.~Acciarri {\it et al.}  (L3 Collaboration),
Phys.\ Lett.\ {\bf B489} (2000) 65
[hep-ex/0005045];\\
H.~Stenzel,
MPI-PHE-99-09
{\it Prepared for 34th Rencontres de Moriond:} ``QCD and Hadronic interactions'',
  Les Arcs, France, 20-27 Mar 1999;\\
%
ALEPH Collaboration,
``QCD Measurements in e+e- Annihilations at Centre-of-Mass 
Energies between 189 and 202\GeV'',
ALEPH 2000-044 CONF 2000-027;



\bibitem{exppowerDIS}  
C.~Adloff {\it et al.}  (H1 Collaboration),
Phys.\ Lett.\ {\bf B406} (1997) 256
[hep-ex/9706002];\\
C.~Adloff {\it et al.}  (H1 Collaboration),
Eur.\ Phys.\ J.\ {\bf C14} (2000) 255
[hep-ex/9912052], erratum {\emph ibid.} {\bf C18} (2000) 417.


\bibitem{Delphi} DELPHI Collaboration, ``The Running of the Strong
  Coupling and a Study of Power Corrections to Hadronic Event Shapes
  with the DELPHI Detector at LEP'', DELPHI 2000-116 CONF 415, July
  2000.

\bibitem{HERApscheme}
G.~J.~McCance  (for the H1 and ZEUS collaborations),
Talk given at \emph{35th Rencontres de Moriond: QCD and High Energy
  Hadronic Interactions} (March 2000),
hep-ex/0008009;\\
H.~Martyn  (for the H1 and ZEUS Collaborations),
Talk given at \emph{30th International Conference on High-Energy Physics}
(ICHEP 2000), hep-ex/0010046.

\bibitem{FeynmanTube}
R.P.~Feynman, `Photon Hadron Interactions', W.A.~Benjamin, New York
(1972).

\bibitem{coherence}
A.~H.~Mueller,
Phys.\ Lett.\ {\bf B104} (1981) 161;\\
B.~I.~Ermolaev and V.~S.~Fadin,
JETP Lett.\ {\bf 33} (1981) 269;\\
Y.~L.~Dokshitzer, V.~S.~Fadin and V.~A.~Khoze,
Z.\ Phys.\ {\bf C15} (1982) 325;\\
A.~Bassetto, M.~Ciafaloni, G.~Marchesini and A.~H.~Mueller,
Nucl.\ Phys.\ {\bf B207} (1982) 189;\\
A.~Bassetto, M.~Ciafaloni and G.~Marchesini,
Phys.\ Rept.\ {\bf 100} (1983) 201.

\bibitem{Herwig}
G.~Marchesini, B.~R.~Webber, G.~Abbiendi, I.~G.~Knowles, M.~H.~Seymour
and L.~Stanco, 
Comput.\ Phys.\ Commun.\  {\bf 67} (1992) 465.


\bibitem{Ariadne}
L.~L\"onnblad,
Comput.\ Phys.\ Commun.\ {\bf 71} (1992) 15; we have use Ariadne
version 4.08 converted to double precision to enable it to be
interfaced to Pythia 6, instead of the older single-precision Jetset;
As a consequence we use identical fragmentation models for Pythia and Ariadne.


\bibitem{Pythia}
T.~Sj\"ostrand,
Comput.\ Phys.\ Commun.\  {\bf 82} (1994) 74;\\
T.~Sj\"ostrand, P.~Eden, C.~Friberg, L.~L\"onnblad, G.~Miu, S.~Mrenna and E.~Norrbin,
hep-ph/0010017.

\bibitem{GBU} Sergio Leone,  \emph{Il Buono, il Brutto, il Cattivo}, 
Italy 1966.

\bibitem{Tkachov} N.~A.~Sveshnikov and F.~V.~Tkachov,
Phys.\ Lett.\ B {\bf 382} (1996) 403
[hep-ph/9512370];\\
P.~S.~Cherzor and N.~A.~Sveshnikov,
presented at \emph{12th International Workshop High-Energy Physics and
  Quantum Field Theory (QFTHEP 97)}, Samara, Russia, September 1997,
hep-ph/9710349;\\
F.~V.~Tkachov,
hep-ph/9901444.

\bibitem{DTProc}
Yu.L. Dokshitzer and S.I. Troyan, \emph{Proceedings of the XIX Winter
  School of the LNPI}, volume 1, p.~144. Leningrad, 1984.

\bibitem{LPHD}
Ya.I. Azimov, Yu.L. Dokshitzer, V.A. Khoze and S.I. Troyan,
Z.\ Phys.\ {\bf C27} (1985) 65.

\bibitem{KLO}
V.~A.~Khoze, S.~Lupia and W.~Ochs,
Eur.\ Phys.\ J.\  {\bf C5} (1998) 77
[hep-ph/9711392].



\bibitem{Basics}
Yu.L. Dokshitzer, V.A. Khoze, A.H. Mueller and S.I. Troyan,
\emph{Basics of Perturbative QCD}, Editions Fronti\`eres, 1991,
Gif-sur-Yvette, France.

\bibitem{MLLA}
A.~H.~Mueller,
Nucl.\ Phys.\  {\bf B213} (1983) 85, erratum quoted \emph{ibid.}, {\bf
  B241} (1984) 141.


\bibitem{tuning}
P.~Abreu {\it et al.}  (DELPHI Collaboration),
Z.\ Phys.\ {\bf C73} (1996) 11.



\bibitem{EEResults}
D.~Decamp {\it et al.}  (ALEPH Collaboration),
Phys.\ Lett.\ {\bf B284} (1992) 163;\\
D.~Buskulic {\it et al.}  (ALEPH Collaboration),
Z.\ Phys.\ {\bf C55} (1992) 209;\\
I.~H.~Park {\it et al.}  (AMY Collaboration),
Phys.\ Rev.\ Lett.\ {\bf 62} (1989) 1713;\\
Y.~K.~Li {\it et al.}  (AMY Collaboration),
Phys.\ Rev.\ {\bf D 41} (1990) 2675;\\
H.~J.~Behrend {\it et al.}  (CELLO Collaboration),
Z.\ Phys.\ {\bf C44} (1989) 63;\\
D.~Bender {\it et al.} (HRS Collaboration),
Phys.\ Rev.\ {\bf D 31} (1985) 1;\\
P.~A.~Movilla Fernandez, O.~Biebel, S.~Bethke, S.~Kluth and P.~Pfeifenschneider
                  (JADE Collaboration),
Eur.\ Phys.\ J.\ {\bf C1} (1998) 461
[hep-ex/9708034];\\
B.~Adeva {\it et al.}  (L3 Collaboration),
Z.\ Phys.\ {\bf C55} (1992) 39;\\
M.~Acciarri {\it et al.}  (L3 Collaboration),
Phys.\ Lett.\ {\bf B411} (1997) 339;\\
D.~P.~Barber {\it et al.} (Mark J Collaboration),
Phys.\ Rev.\ Lett.\ {\bf 43} (1979) 830;\\
A.~Petersen {\it et al.} (Mark II Collaboration),
Phys.\ Rev.\ {\bf D 37} (1988) 1;\\
S.~Bethke {\it et al.} (Mark II Collaboration),
Z.\ Phys.\ {\bf C43} (1989) 325;\\
P.~D.~Acton {\it et al.}  (OPAL Collaboration),
Z.\ Phys.\ {\bf C59} (1993) 1;\\
G.~Alexander {\it et al.}  (OPAL Collaboration),
Z.\ Phys.\ {\bf C72} (1996) 191;\\
K.~Ackerstaff {\it et al.}  (OPAL Collaboration),
Z.\ Phys.\ {\bf C75} (1997) 193;\\
C.~Berger {\it et al.}  (PLUTO Collaboration),
Z.\ Phys.\ {\bf C12} (1982) 297;\\
K.~Abe {\it et al.}  (SLD Collaboration),
Phys.\ Rev.\ {\bf D 51} (1995) 962
[hep-ex/9501003];\\
W.~Braunschweig {\it et al.} (TASSO Collaboration),
Phys.\ Lett.\ {\bf B214} (1988) 286;\\
W.~Braunschweig {\it et al.}  (TASSO Collaboration),
Z.\ Phys.\ {\bf C45} (1989) 11;\\
W.~Braunschweig {\it et al.}  (TASSO Collaboration),
Z.\ Phys.\ {\bf C47} (1990) 187;\\
I.~Adachi {\it et al.}  (TOPAZ Collaboration),
Phys.\ Lett.\ {\bf B227} (1989) 495;\\
K.~Nagai {\it et al.}  (TOPAZ Collaboration),
Phys.\ Lett.\ {\bf B278} (1992) 506;\\
Y.~Ohnishi {\it et al.}  (TOPAZ Collaboration),
Phys.\ Lett.\ {\bf B313} (1993) 475.




\bibitem{Nason:1994xx}
P.~Nason and B.~R.~Webber,
Nucl.\ Phys.\ {\bf B421} (1994) 473, erratum \emph{ibid.} {\bf B480}
755 (1996). 

\bibitem{Nason:1998nw}
P.~Nason and C.~Oleari,
Nucl.\ Phys.\ {\bf B521} (1998) 237
[hep-ph/9709360].

\bibitem{Oleari:1997az}
C.~Oleari, Ph.D. thesis,
hep-ph/9802431.

\bibitem{Rodrigo:1997ha}
G.~Rodrigo,
Nucl.\ Phys.\ Proc.\ Suppl.\ {\bf 54A} (1997) 60
[hep-ph/9609213].

\bibitem{Rodrigo:1996gw}
G.~V.~Rodrigo, Ph.D. thesis,
hep-ph/9703359.

\bibitem{Rodrigo:1997gy}
G.~Rodrigo, A.~Santamaria and M.~Bilenkii,
Phys.\ Rev.\ Lett.\ {\bf 79} (1997) 193
[hep-ph/9703358].

\bibitem{Bernreuther:1997jn}
W.~Bernreuther, A.~Brandenburg and P.~Uwer,
Phys.\ Rev.\ Lett.\ {\bf 79} (1997) 189
[hep-ph/9703305].

\bibitem{Brandenburg:1998pu}
A.~Brandenburg and P.~Uwer,
Nucl.\ Phys.\ {\bf B515} (1998) 279
[hep-ph/9708350].


\bibitem{Abbiendi:1999fs}
G.~Abbiendi {\it et al.}  (OPAL Collaboration),
Eur.\ Phys.\ J.\ {\bf C11} (1999) 643
[hep-ex/9904013].



\bibitem{Catani:1998sf}
S.~Catani and B.~R.~Webber,
Phys.\ Lett.\  {\bf B427} (1998) 377
[hep-ph/9801350].



\bibitem{SZ}
G.~P.~Salam and G.~Zanderighi,
Nucl.\ Phys.\ Proc.\ Suppl.\  {\bf 86} (2000) 430
[hep-ph/9909324].


\end{thebibliography}
\end{document}